\documentclass[preprint,aps,pre,floatfix]{revtex4-1}

\usepackage{graphicx}
\usepackage{epstopdf} 
\usepackage{dcolumn}
\usepackage{bm}
\usepackage{color}

\begin{document}

%
%

\title{Kinetic Simulations of Plasmoid Chain Dynamics}

\author{S. Markidis}
\email{markidis@pdc.kth.se}
\affiliation{High Performance Computing and Visualization (HPCViz) Department, KTH Royal Institute of Technology, Stockholm, Sweden}
\author{P. Henri}
\affiliation{Universit\'e de Nice Sophia Antipolis, CNRS, Observatoire de la C\^ote d'Azur, Nice, France}
\author{G. Lapenta}
\affiliation{Centrum voor Plasma-Astrofysica, 
Department Wiskunde, Katholieke Universiteit Leuven,  Leuven, Belgium}
\author{A. Divin}
\affiliation{Swedish Institute of Space Physics, Uppsala, Sweden}
\author{M. Goldman}
\author{D. Newman}
\affiliation{Department of Physics and CIPS, University of Colorado, Boulder, USA}
\author{E. Laure}
\affiliation{PDC and High Performance Computing and Visualization (HPCViz) Department, KTH Royal Institute of Technology, Stockholm, Sweden}

\date{\today}

\begin{abstract}
The dynamics of a plasmoid chain is studied with three dimensional Particle-in-Cell simulations. The evolution of the system with and without a uniform guide field, whose strength is 1/3 the asymptotic magnetic field, is investigated. The plasmoid chain forms by spontaneous magnetic reconnection: the tearing instability rapidly disrupts the initial current sheet generating several small-scale plasmoids, that rapidly grow in size coalescing and kinking. The plasmoid kink is mainly driven by the coalescence process. It is found that the presence of guide field strongly influences the evolution of the plasmoid chain. Without a guide field, a main reconnection site dominates and smaller reconnection regions are included in larger ones, leading to an hierarchical structure of the plasmoid-dominated current sheet. On the contrary in presence of a guide field, plasmoids have approximately the same size and the hierarchical structure does not emerge, a strong core magnetic field develops in the center of the plasmoid in the direction of the existing guide field, and bump-on-tail instability, leading to the formation of electron holes, is detected in proximity of the plasmoids. 
\end{abstract}
\maketitle

%
%

\section{Introduction}

Magnetic reconnection is an ubiquitous phenomenon in space, astrophysical and laboratory plasmas causing the conversion of magnetic field energy into kinetic energy of plasmas by a reorganization of the magnetic field topology. Magnetic reconnection occurs in localized spatial region, called $x$ lines, where plasma is accelerated into/out. When magnetic reconnection develops in multiple $x$ lines \citep{Hughes:1987, Slavin:2003}, the outflow plasmas from neighbor reconnection sites form high density structures, the so-called "plasmoids" \citep{Hones:1977}, organized as beads in a chain. Plasmoid chains are highly dynamic configurations: plasmoids can rapidly grow by coalescence \citep{Bhattacharjee:1983}, bounce \citep{Intrator:2009,Karimabadi:2011} and eject smaller plasmoids \citep{Daughton:2009}. Observational studies revealed the presence of  multiple plasmoids in planetary magnetospheric environments \citep{Lin:2008, Khotyaintsev:2010,Borg:2012,slavin2012messenger}. In addition, the possible presence of plasmoid chains has been suggested in the solar corona \citep{Bemporad:2008}. It has also been shown that the small scale dynamics of plasmoid chains strongly influences the large scale, fluid, dynamics of magnetized plasmas, for instance during the non-linear evolution of shear flows, where the non-linear dynamics of a chain of magnetic islands (the 2D equivalent of plasmoids) forming inside Kelvin-Helmholtz vortices can disrupt these vortices and prevent their pairing \citep{Henri:2012}.

Several recent theoretical and computational investigations focus on plasmoid chain dynamics. Early simulations studied the occurrence of plasmoid chain by MHD simulations \citep{Biskamp:1986,Malara:1992}. Theoretical models and simulations showed how the occurrence of secondary plasmoids affects magnetic reconnection and how fast reconnection in low resistivity plasmas can be achieved in the MHD framework via stochastic magnetic reconnection \citep{Lazarian:1999, Kowal:2009, Loureiro:2005, Loureiro:2007, Lapenta:2008,bhattacharjee:2009,Cassak:2009, Uzdensky:2010}.  Collisionless Particle-in-Cell simulations \citep{Tajima:1987,Pritchett:2008,OkaAPJ:2010, OkaJGR:2010,Tanaka:2010, Markidis:2012} studied the dynamics of plasmoid chain focusing on kinetic effects and acceleration mechanisms in multiple magnetic island magnetic reconnection \citep{Hoshino:2012, Drake:Nature}. Previous particle simulation studies used a reduced two dimensional geometry, neglecting current-aligned instabilities developing in the third direction, or modeled pair relativistic plasmas in three dimensions \citep{Kagan:2012}. Differently from previous works on collisionless plasmoid chain, Particle-in-Cell simulations in three dimension with particle with higher mass ratios than previous studies and in a simulation set-up that resemble the Earth's magnetotail are presented in this article. This study shows that the presence of a guide field, an initial magnetic field with direction perpendicular to the reconnection plane, strongly affects the dynamics of plasmoid by changing the overall structure of the plasmoid chain and introducing different instabilities in the system. To understand the role of guide field, two Particle-in-Cell simulations are carried out starting from a magnetic field configuration with and without a guide field.

The paper is organized as follows. First, the simulation model is presented in Section II. The formation and evolution of plasmoid chain in antiparallel configuration is analyzed by studying the evolution of densities, magnetic and electric fields, flow patterns and distribution functions in Section III. The same analysis is repeated for the simulations starting from a configuration with guide field and presented in Section IV. The results of the two simulations are discussed and compared with the results of previous work in Section V. Finally, the main results are summarized in Section VI.
\section{Simulation Parameters}
Simulations are carried out in a three-dimensional system, where an Harris current sheet configuration is initially imposed. The $z$ coordinate is taken along the Harris sheet current, while the $x$-$y$ plane is the reconnection plane. 
The plasma density profile is initialized as:  
\begin{equation}
n(y) = 0.2 \ n_0 \cosh^{-2}\left(\frac{y - L_y/2}{\lambda}\right)  + n_{b}
\end{equation}
The peak density $n_0$ is the reference density, while the background density $n_{b} = 0.2\ n_0$; $L_x \times L_y  \times L_z = 40 \ d_i \times 15 \ d_i \times 10 \ d_i$ are the simulation box lengths and $\lambda = 0.5 \ d_i$ is the half-width of the current sheet. The ion inertial length is $d_i = c/\omega_{pi}$,  with $c$ the speed of light in vacuum, $\omega_{pi} = \sqrt{4 \pi n_0e^2/m_i}$ the ion plasma frequency, $e$ the elementary charge, and $m_i$ the ion mass. The  electron mass is $m_e = m_i/256$. 
A magnetic field in the $x$ direction and varying in the $y$ direction, is initialized:
 \begin{equation}
B_x(y)=B_0\tanh\left(\frac{y - L_y/2}{\lambda}\right).
\end{equation}
Two simulations are carried out: a first simulation is completed starting from the initial condition described above, while a uniform guide field $B_g = 1/3 \ B_0$ along the $z$ direction is added to the magnetic field configuration in the second simulation. This guide field value has been observed during magnetic reconnection in the Earth's magnetotail \citep{Wang:2012}. These two initial configurations has been selected to show how a relatively weak guide field can dramatically change the dynamics of plasmoid chain.

In these simulations, spontaneous magnetic reconnection is triggered following the approach presented in Refs.~\citep{Newman:2012, Markidis:2012}: the current, supporting the initial magnetic field configuration, is not present. As a result of this non-equilibrium, the plasma is initially accelerated toward the current sheet to establish a current consistent with the magnetic field configuration, collapsing the current sheet, triggering the tearing instability and forming plasmoid structures with spatial scales approximately $1\ d_i$ large. This non-equilibrium initial condition accelerates the reconnection initiation and the following coalescence process of the plasmoids. This approach is different from the initial condition used in previous two-dimensional Particle-in-Cell simulations~\citep{Pritchett:2007,Karimabadi:2011}. These studies used the island-chain equilibrium, or Fadeev equilibrium~\citep{Fadeev:1965}, that imposes the initial presence of spatially larger scale magnetic islands where the coalescence process involves larger scale plasmoids and proceeds at a slower pace with respect to our simulations.

The particles are initialized with a Maxwellian velocity distribution. The electron thermal velocity $v_{the}/c = (T_e/(m_e c^2))^{1/2} =  0.045$, while the ion temperature $T_i = 5 \ T_e$. This temperature ratio is typical of the Earth's magnetotail. The simulation time step is $\omega_{pi}\Delta t = \ 0.125$. In this simulation set-up, $\omega_{pi}/\Omega_{ci} = c/ V_A = 103$, where $\Omega_{ci} = e B_0/m_i$ is the ion cyclofrequency and $V_{A}$ is the Alfv\'en velocity calculated with $B= B_0$ and $n = n_0$. The ion Larmor gyro-radius is $0.65 \ d_i$. 

The grid is composed of $512 \times 192 \times 128$ cells. In total $3\times 10^9$ computational particles are in use. The boundaries are periodic in the $x$ and $z$ directions, while they are perfect conductor and reflecting boundary conditions for fields and particles respectively in the $y$ direction. Simulations are carried out with the massively parallel implicit Particle-in-Cell {\em iPIC3D} code~\citep{Markidis:2010}, running on 12,288 cores on NASA Pleiades supercomputer.
\section{Plasmoid Chain Dynamics in Antiparallel Magnetic Reconnection}
\begin{figure}[ht]
\includegraphics[width=0.75 \columnwidth]{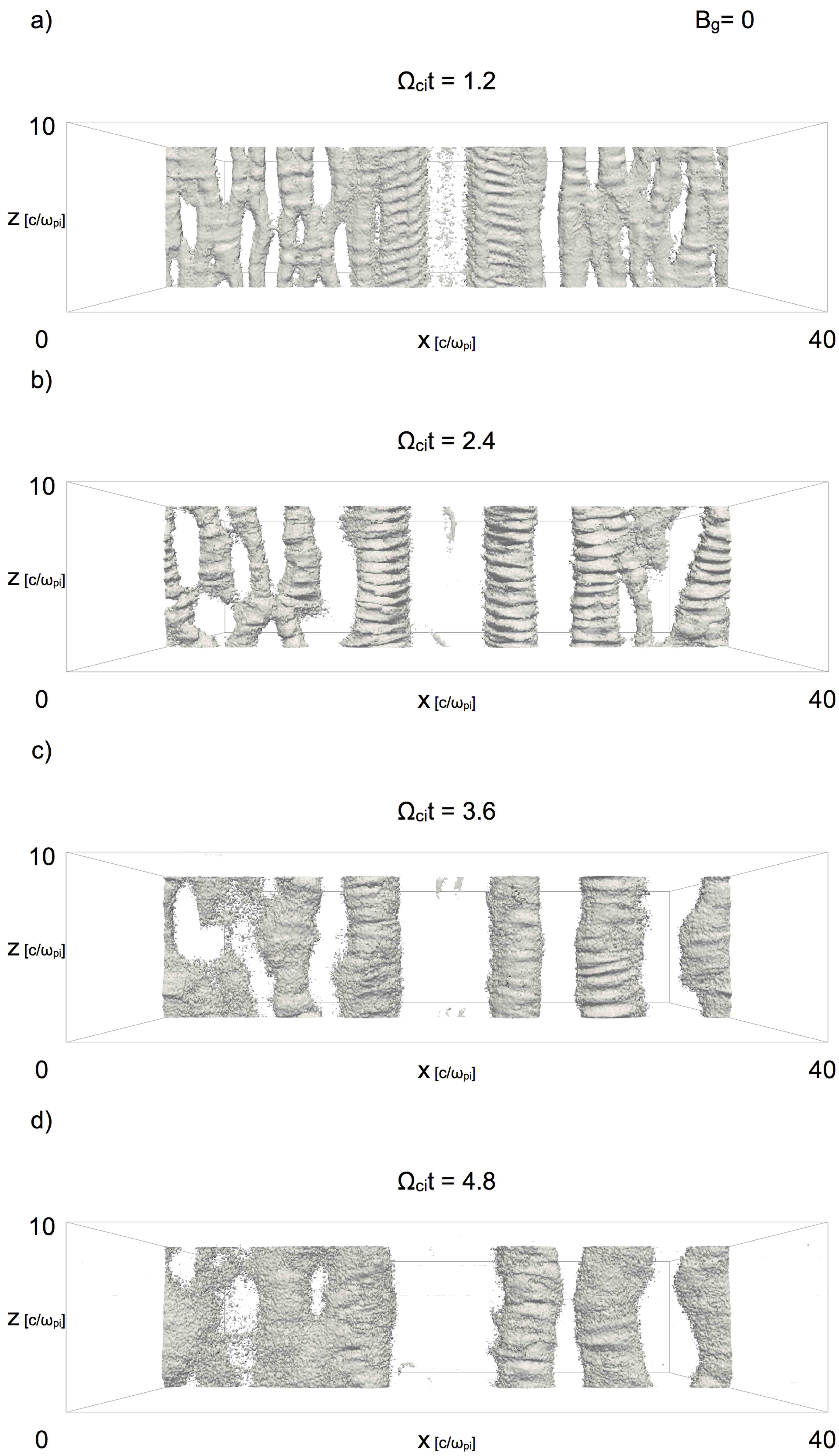}
\caption{Electron density isosurfaces for $n_e = 1.7 \ n_b$ at four different times. Plasmoids appear as high density tubes merging and kinking.}
\label{evolutionNeNOGF}
\end{figure}

\begin{figure}[ht]
\includegraphics[width=0.95 \columnwidth]{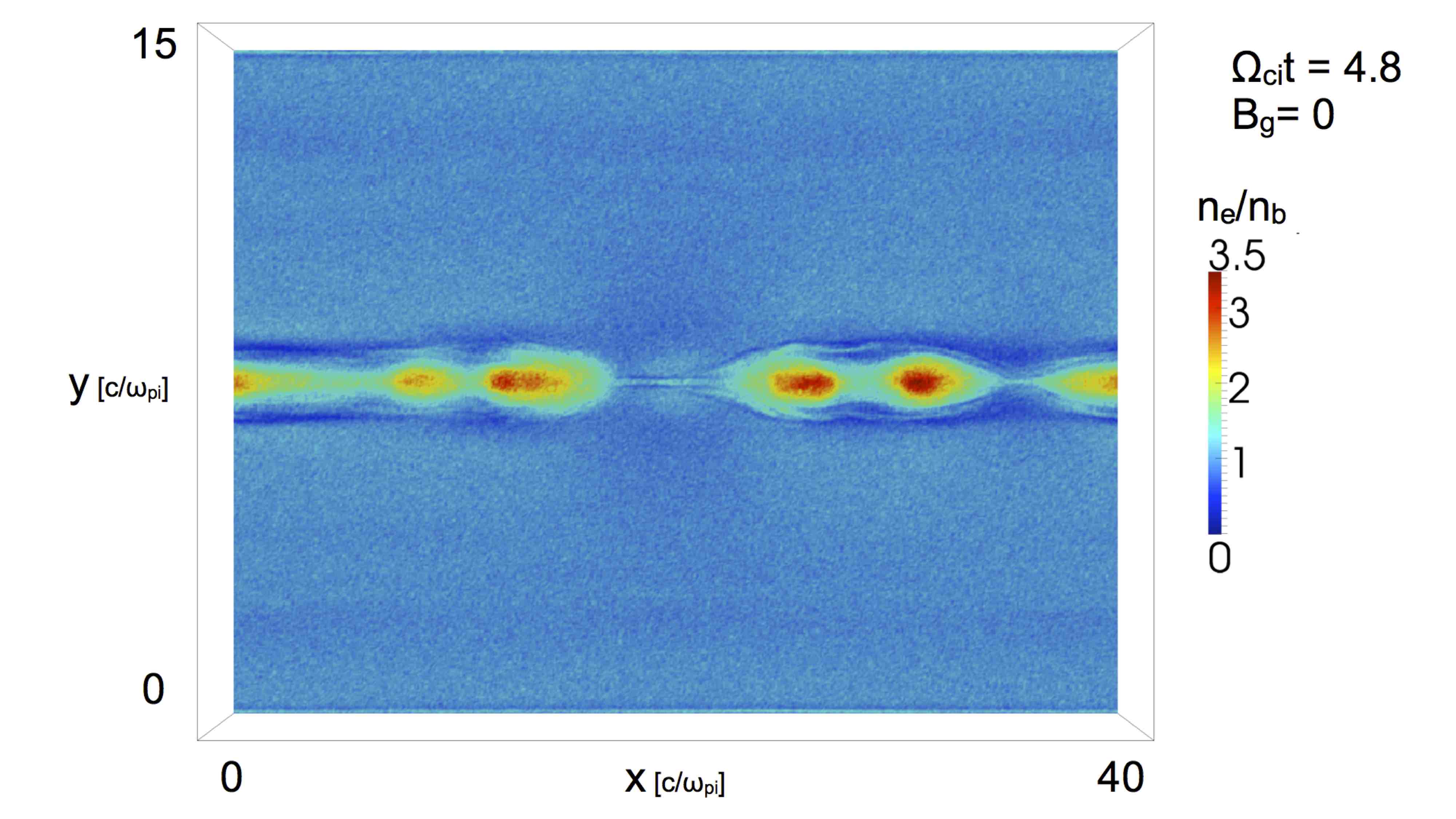}
\caption{Electron density contour-plot at $z = L_z/2$ at time $\Omega_{ci} t = 4.8$. The plot shows the formation a main reconnection site approximately at the center of contour-plot.}
\label{neNOGF}
\end{figure}

\begin{figure}[ht]
\includegraphics[width=0.95 \columnwidth]{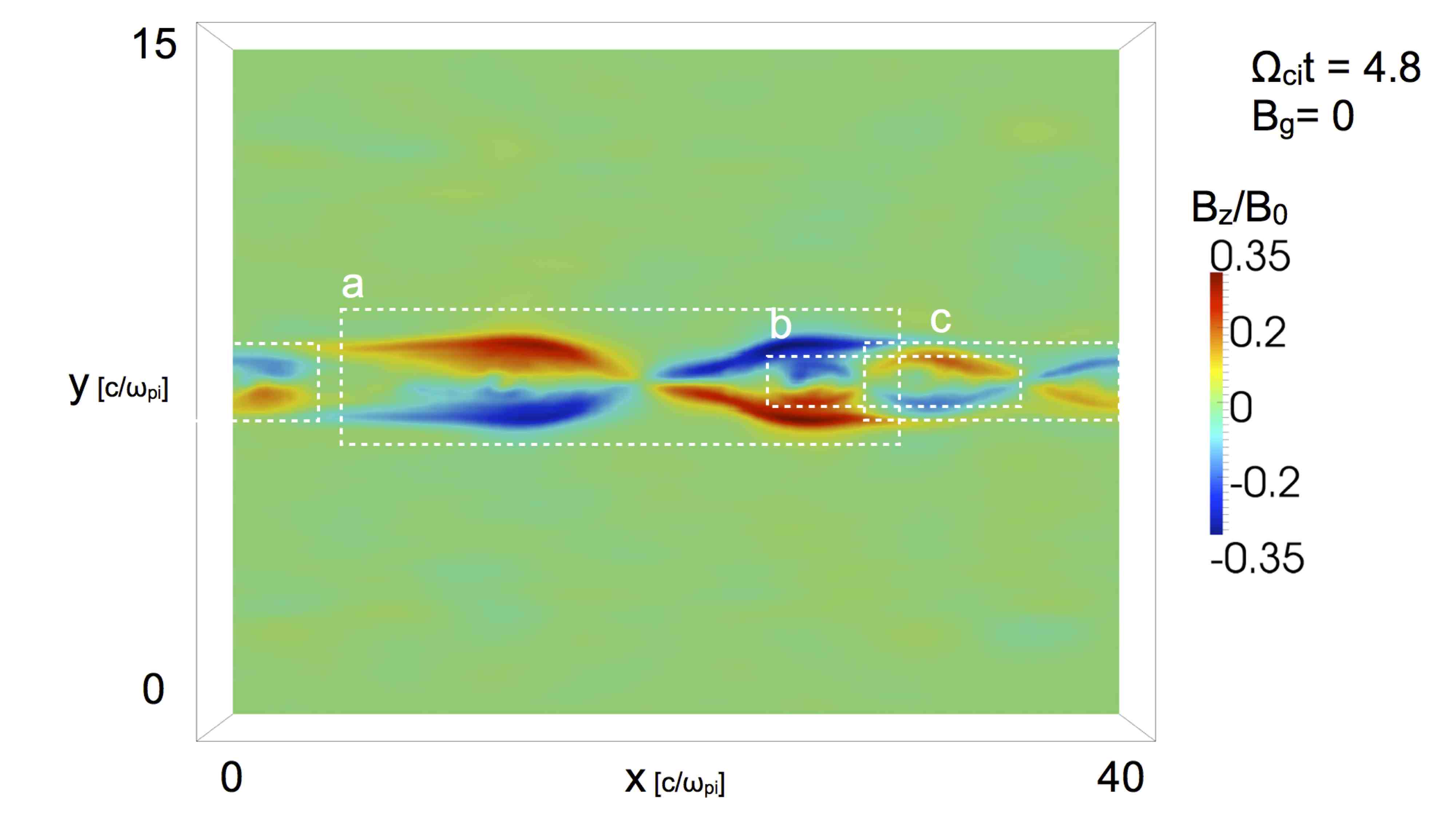}
\caption{Contour-plot of $B_z$ magnetic field component on the $z = L_z/2$ plane at time $\Omega_{ci} t = 4.8$. The dashed line boxes enclose Hall magnetic field structures defining reconnection regions. It is clear that some reconnection regions are enclosed in larger reconnection areas, leading to a hierarchical structure of the plasmoid-dominated current sheet.}
\label{HallMagneticField}
\end{figure}

\begin{figure}[ht]
\includegraphics[width=0.95 \columnwidth]{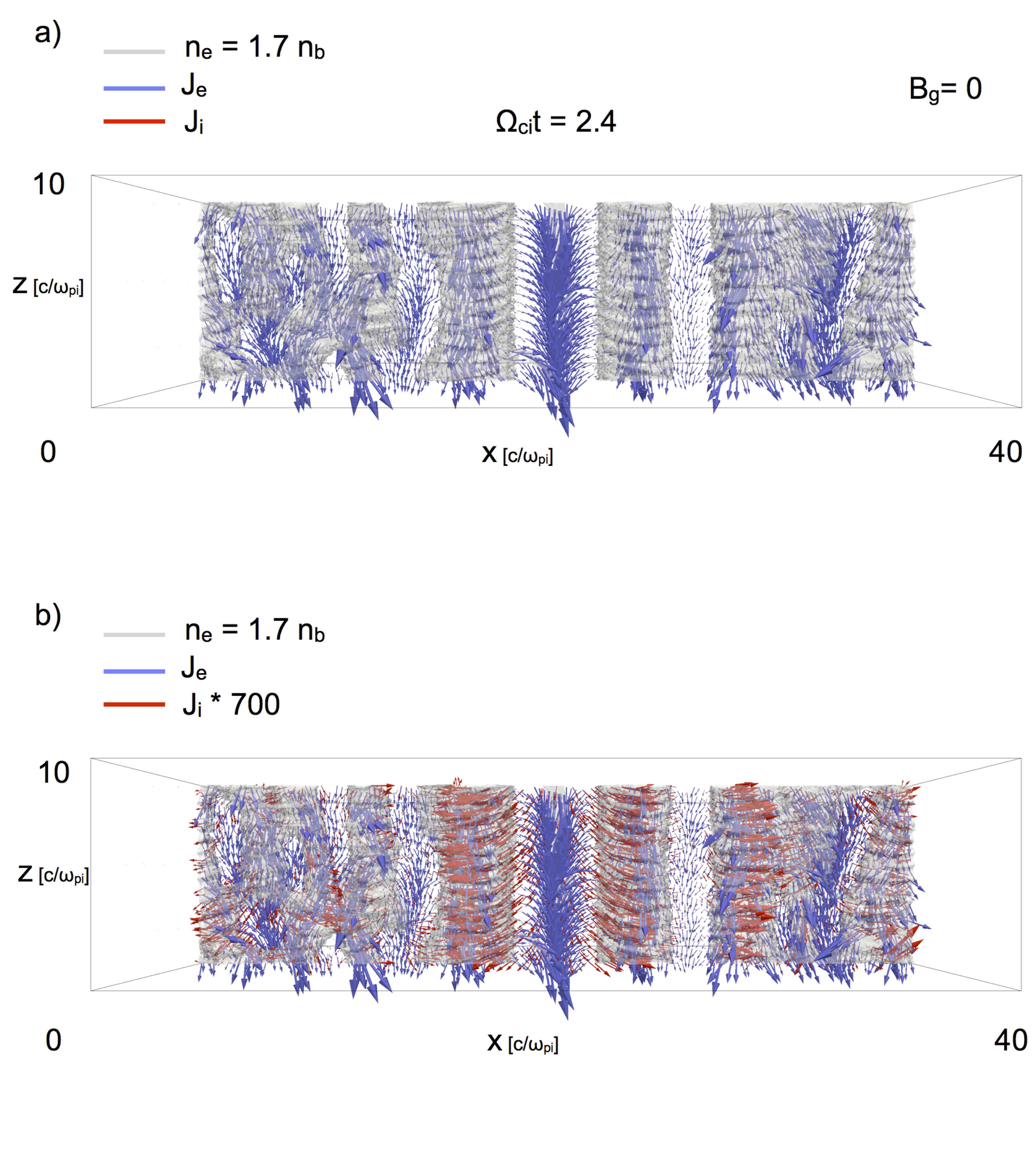}
\caption{Quiver plot of electron (blue color) and ion (red color) currents superimposed to the electron density iso-surface (grey color) for $n_e = 1.7 \ n_b$ in panel a). In panel b) the intensity of the ion current is multiplied by 700 to show the direction of the ion current. The magnetic field loops are supported by an electron current along the $z$ direction, while the ion current (approximately 700 times smaller) develops on the $x-y$ plane to support the Hall magnetic field.}
\label{J_NOGF}
\end{figure}

\begin{figure}[ht]
\includegraphics[width=0.95 \columnwidth]{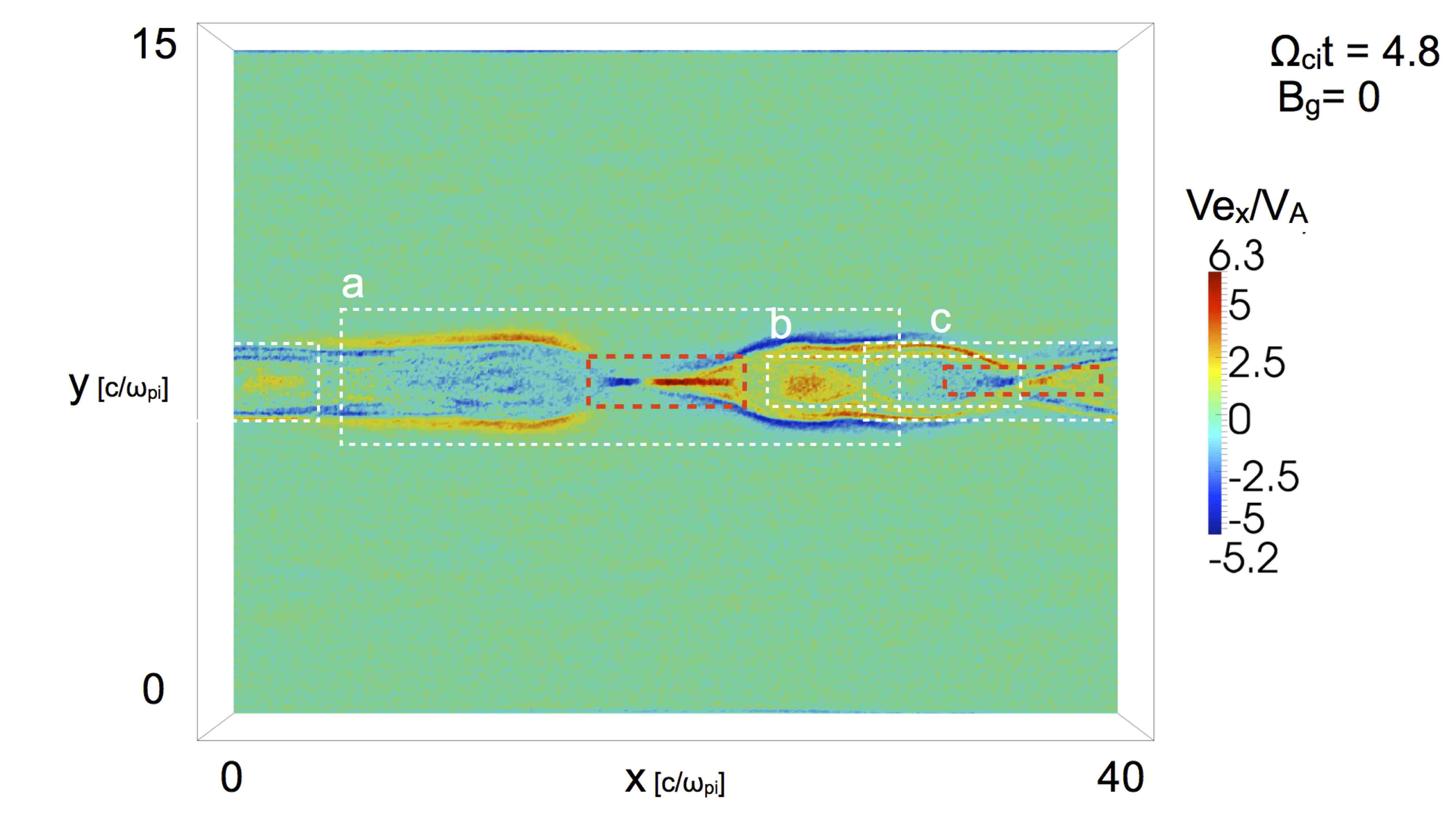}
\caption{Contour-plot of the $x$ component of the electron fluid velocity component on the $z = L_z/2$ plane at time $\Omega_{ci} t = 4.8$. The white dashed line boxes surround the previously identified reconnection regions, while the red dashed line boxes enclose the two electron jets regions.}
\label{VexNOGF}
\end{figure}

A first simulation is carried out, starting from a configuration without a guide field. A tearing instability rapidly occurs and generates several plasmoids. These appear as cylindrical surfaces, characterized by enhanced density with respect to the background density, as clear in Figure \ref{evolutionNeNOGF}, where the electron density isosurfaces for $n_e = 1.7 \  n_b$ are presented at four different times. The plasmoids merge in single points kinking during the coalescence. The kink of the plamoid is not only a consequence of a Kruskal-Shafranov-type kink instability~\citep{Poedts:2004} but it is mainly due to non uniform coalescence process. The electron density isosurfaces appear perturbed by wiggles, denoting the presence of lower hybrid waves in proximity of density gradient regions and propagating along the $z$ direction~\citep{yoon:2002,Daughton:2003}. These perturbations are stronger earlier in time (panel a and b). A rather large low density region develops approximately at the center of the simulation box at time $\Omega_{ci}t = 4.8$.
This can be easily seen in Figure \ref{neNOGF}, where a contour-plot of the electron density on the $z = L_z/2$ plane at time $\Omega_{ci} t = 4.8$ is presented. At time $\Omega_{ci}t = 4.8$, five plasmoids are visible in the figure. Their electron density is 3.5 times larger than the background density, and they are approximately $5 \ d_i$ large. 
The tearing instability leads to the formation of magnetic field loops surrounding the plasmoids as clear in Figure \ref{neNOGF}. The magnetic field lines are on the $x-y$ plane with no strong component in the $z$ direction.
\begin{figure}[ht]
\includegraphics[width=0.95 \columnwidth]{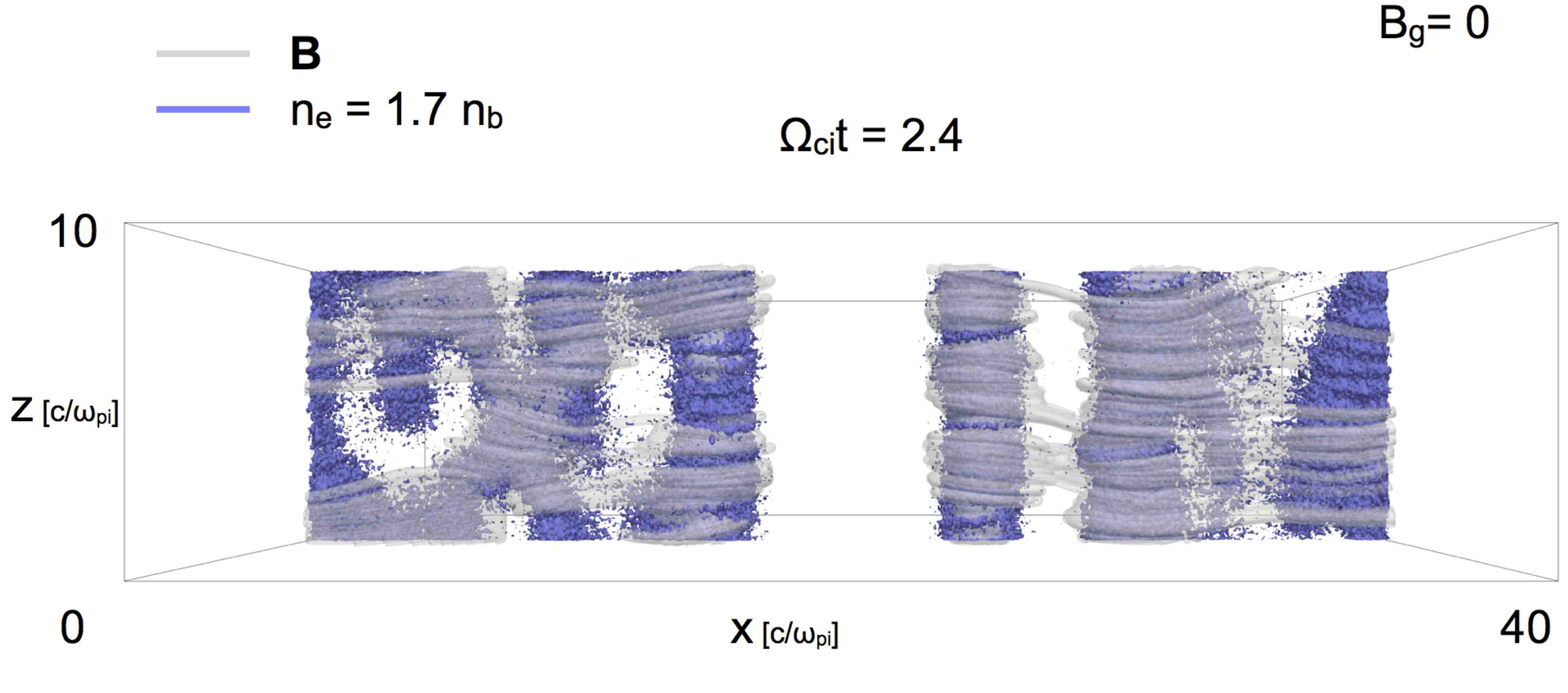}
\caption{Magnetic field lines in grey color superimposed to the electron density iso-surface in blue color at time $\Omega_{ci} t = 4.8$. The magnetic field loops surround the plasmoids.}
\label{BlinesNOGF}
\end{figure}
However, the $B_z$ quadrupolar structure, typical of Hall magnetic reconnection, is present and visible in Figure \ref{HallMagneticField}. The dashed line boxes in this figure surround three magnetic reconnection areas. The reconnection region $b$ is enclosed in the reconnection region $a$, while only a part of the reconnection region $c$ is surrounded by the region $a$.
It is clear from an inspection of the current system in Figure \ref{J_NOGF} that the magnetic field loops (shown in Figure \ref{BlinesNOGF}) are only supported by the electron current, while ion dynamics develops in the $x-y$ plane. The decoupling of electron and ion dynamic on this plane leads to the formation of the Hall magnetic field. Panel a) of Figure \ref{J_NOGF} shows a quiver plot of the electron and ion currents in blue and red colors superimposed to electron density isosurfaces. Ion currents are very weak if compared to electron current and they are not visible in the plot. In panel b) the intensity of the ion current is scaled 700 times to show the direction of the ion current.
Electron jets emanate from the $x$ lines and form the outer electron diffusion regions in antiparallel reconnection. They have been identified in previous studies of collisionless reconnection ~\citep{Nagai:2001,karimabadi:2007, Nagai:2013}  and are detected in this work also. Figure \ref{VexNOGF} shows a contour-plot of the $x$ component of the electron fluid velocity ($\mathbf{v}_e = \mathbf{J}_e/\rho_e$). Multiple electron jets exit the $x$ lines and they are shown in the dashed line boxes in Figure \ref{VexNOGF}. 
\section{Plasmoid Chain Dynamics in Presence of a Guide Field}
\begin{figure}[ht]
\includegraphics[width=0.75 \columnwidth]{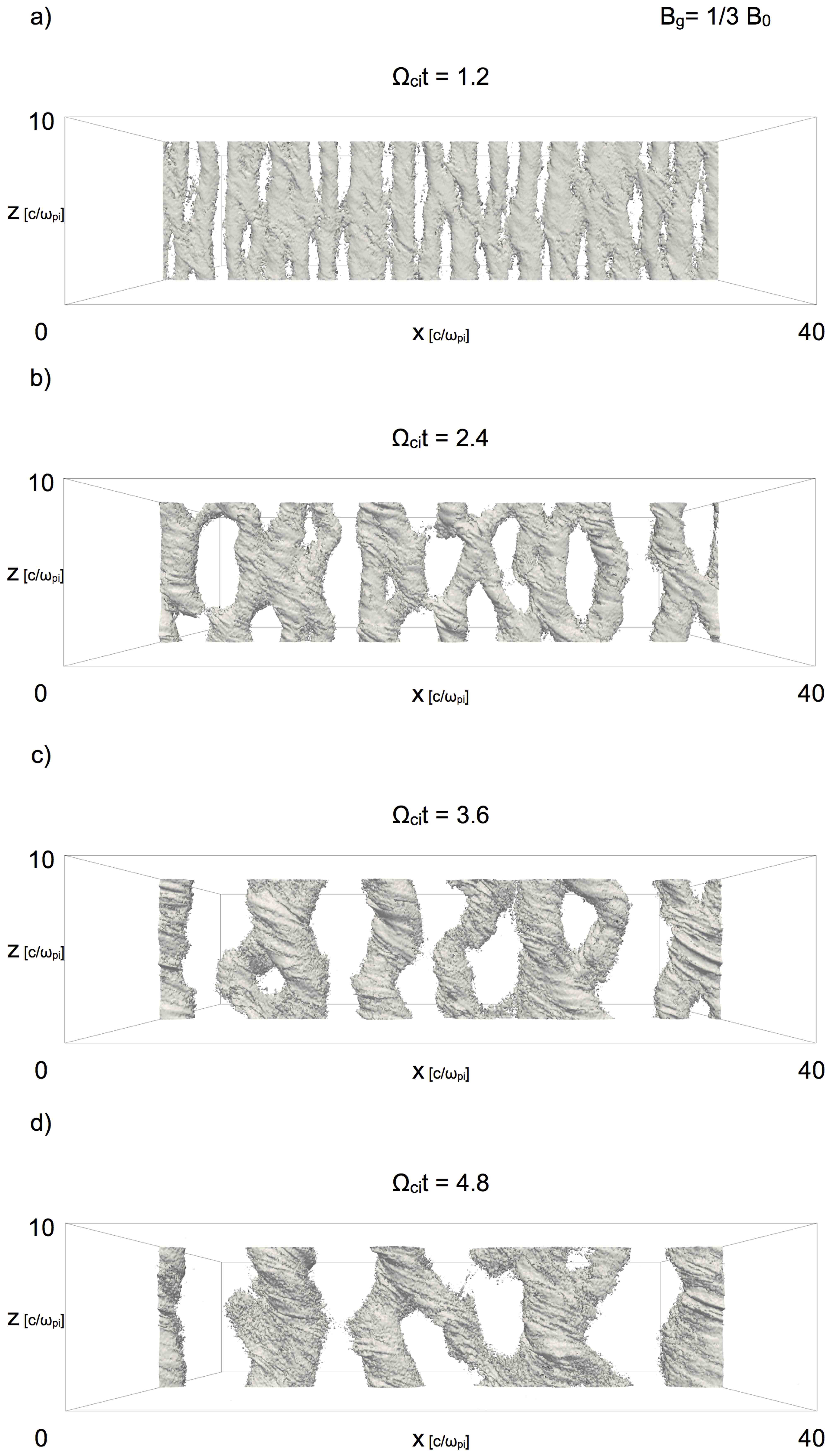}
\caption{Electron density isosurfaces for $n_e = 1.7 \ n_b$ at four different times. Plasmoids appear has high density tubes merging and kinking as in the antiparallel set-up, but in addition the plasmoid tubes twist also.}
\label{neEvolution_GF}
\end{figure}

\begin{figure}[ht]
\includegraphics[width=0.95 \columnwidth]{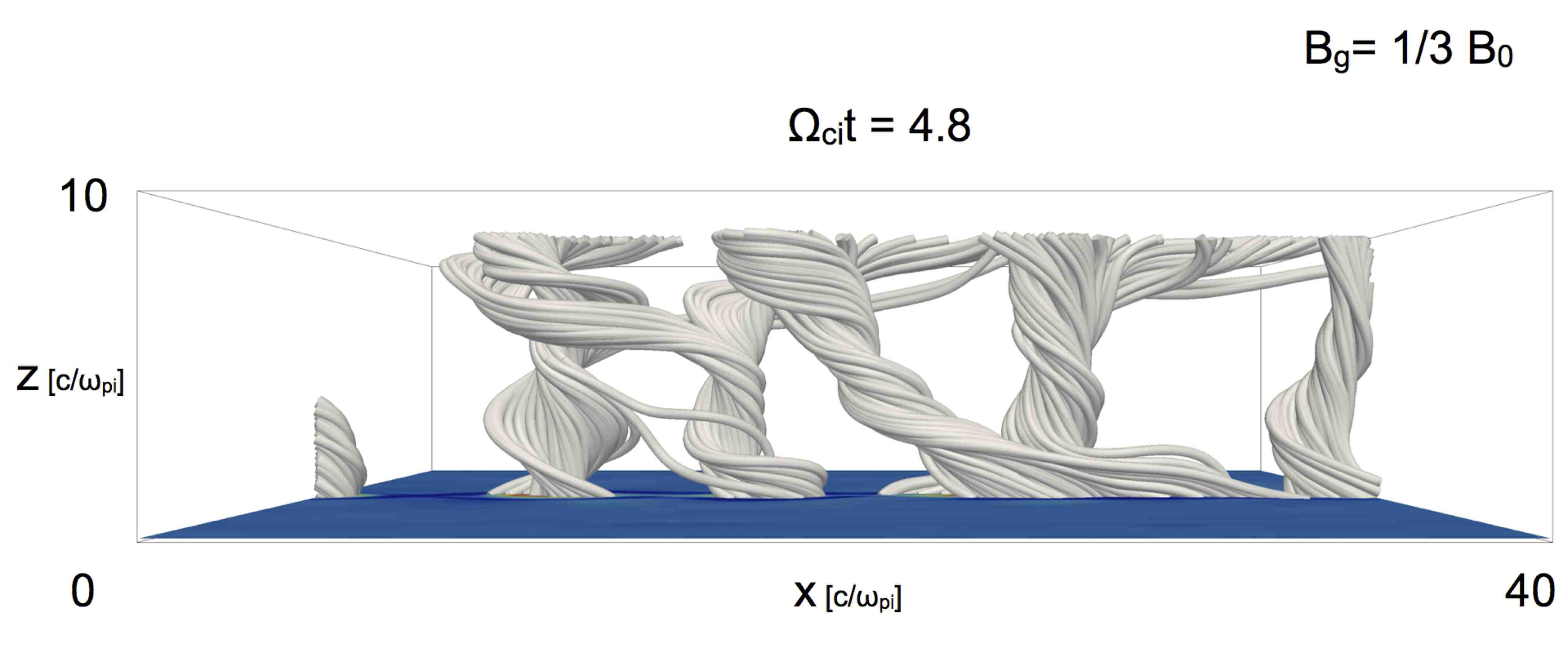} 
\caption{Magnetic field lines at time $\Omega_{ci} t = 4.8$}
\label{B_GF}
\end{figure}

\begin{figure}[ht]
\includegraphics[width=0.95 \columnwidth]{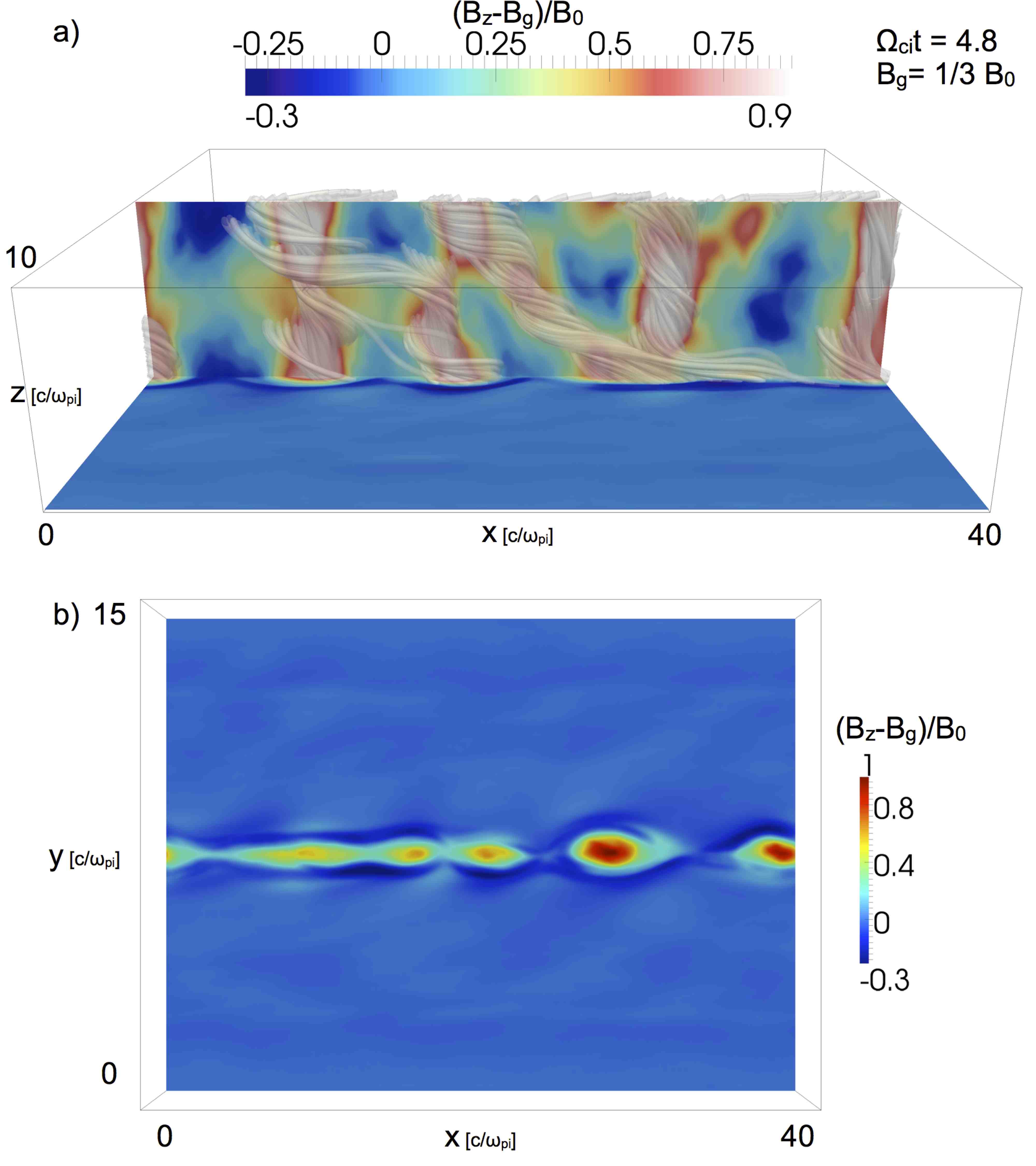}
\caption{Magnetic field lines and contourplot of $(B_z - B_0)/B_0$ on the $y = L_y/2$ plane at time $\Omega_{ci} t = 4.8$ in panel a). A contour-plot of  of $(B_z - B_0)/B_0$ on the $z= L_z/2$ plane. The core field has approximately the same strength and the same direction of the guide field for all the plasmoids.}
\label{Bcore_GF}
\end{figure}

\begin{figure}[ht]
\includegraphics[width=0.95 \columnwidth]{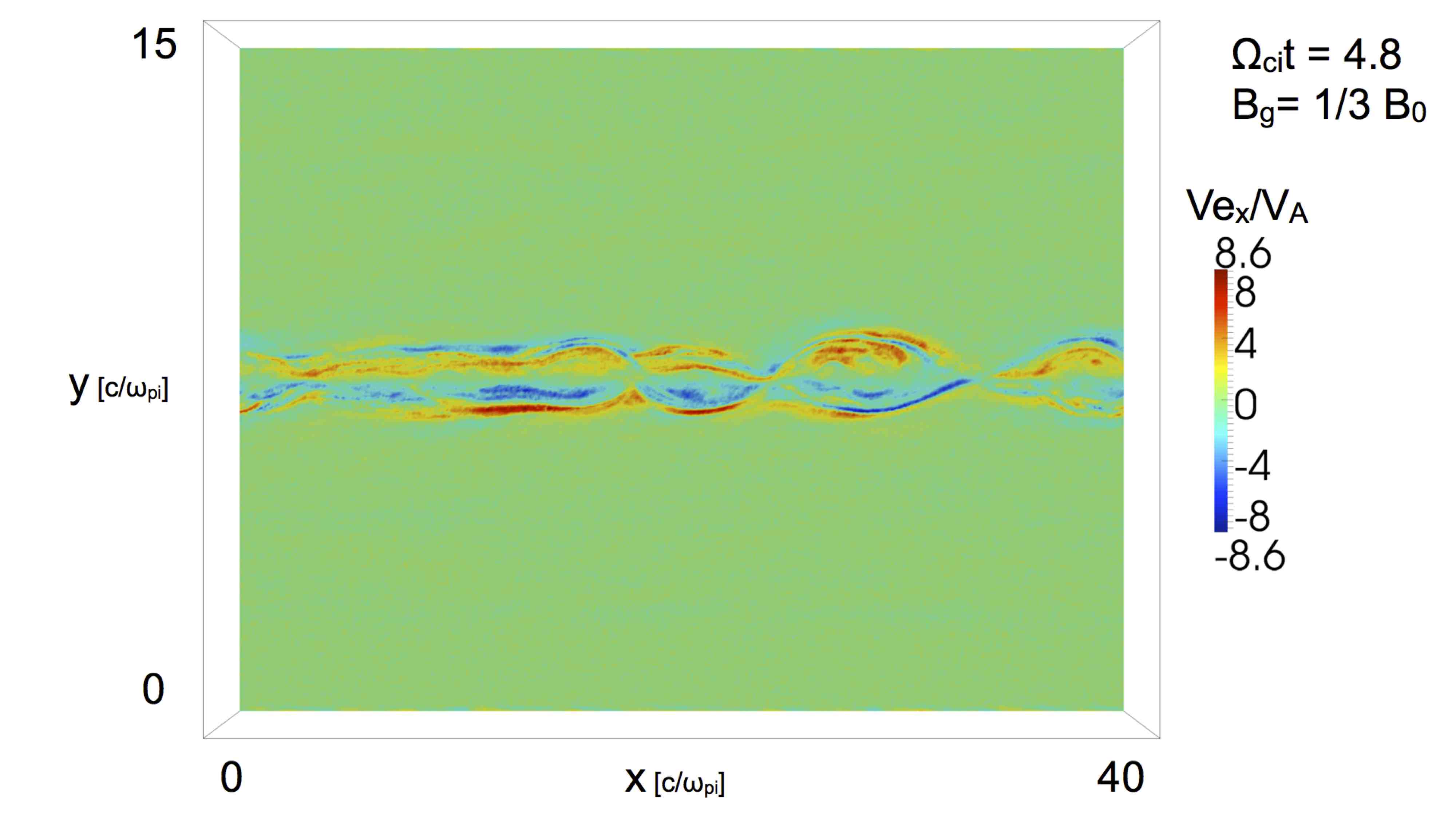}
\caption{Contour-plot of the $x$ component of the electron fluid velocity component on the $z = L_z/2$ plane at time $\Omega_{ci} t = 4.8$. }
\label{VexGF}
\end{figure}

A second simulation has been carried out, starting from a configuration with guide field. Its presence strongly influences the dynamics of the plasmoid chain. As in the first simulation, plasmoids form as a result of the tearing instability. The coalescence process proceeds at a pace that is comparable to the one in the antiparallel case. This is clear from comparing the number of plasmoids present at different snapshots of electron density in Figure \ref{evolutionNeNOGF} and \ref{neEvolution_GF}, where approximately the same number of plasmoids is present at the same time snapshot. In addition, the guide field introduces a twist in the plasmoid structures as macroscopic result of the Lorentz force. This can be seen in panels c) and d) of Figure \ref{neEvolution_GF}. Lower hybrid waves perturb the density isosurfaces, as in the other simulation without a guide field, but propagating obliquely to the guide field and density gradient directions \citep{Daughton:2003}. 
An helix shape magnetic field develops around the plasmoids in the case of guide field reconnection, as visible in the two panels of Figure \ref{B_GF}. We recall that magnetic helicity, that is a conserved quantity in MHD framework, is not necessarily conserved in the kinetic approach.
\begin{figure}[ht]
\includegraphics[width=0.95 \columnwidth]{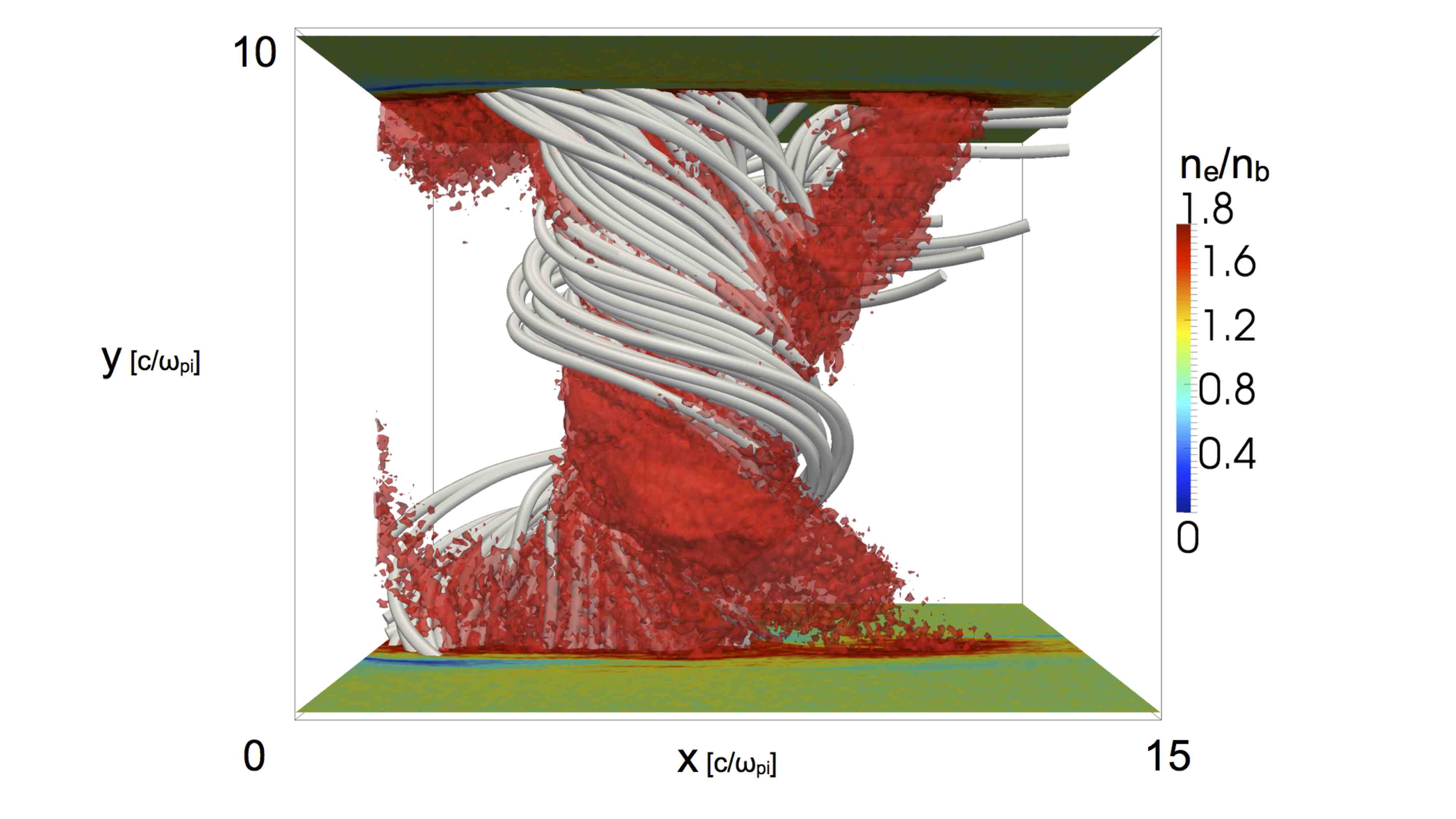}
\caption{A blow-up of the region enclosing a flux rope, showing the magnetic field lines, superimposed to electron density isosurfaces in red color.}
\label{B_GF_2}
\end{figure}
A strong unipolar magnetic field in the $z$ direction forms at the center of the plasmoids.  From panel b) of Figure \ref{Bcore_GF}, where a contour-plot of  of $(B_z - B_0)/B_0$ on the $z= L_z/2$ plane at time $\Omega_{ci} t = 4.8$ is presented, it is clear the core magnetic $B_z - B_g$ field has a value that is approximately equal to the asymptotic magnetic field $B_0$ and it is unipolar.
The outer electron diffusion region is disrupted in guide field reconnection as previous studies showed \citep{Goldman:2011,Lapenta:2010}. Figure \ref{VexGF} shows a contourplot of the $x$ component of the electron fluid velocity, where the electron beams are localized along the separatrices.

In a previous work~\citep{Markidis:2012}, it has been shown that electron streaming instabilities occur during multiple island reconnection in two dimensional geometry. In particular, the electron two-stream instability is shown to be responsible for the formation of bipolar parallel electric field structures surrounding the magnetic islands. Such kinetic instability appears when counter-streaming electrons are found at the same position. In a two dimensional system, this instability cannot develop in the third dimension. It is thus a priori not clear that the same mechanism can be found in tridimensional cases. In this paper, we show that the formation of bipolar electric field structures due to electron streaming kinetic instabilities can also occur in the three dimensional configuration in presence of a guide field.
\begin{figure}[ht]
\includegraphics[width=0.95 \columnwidth]{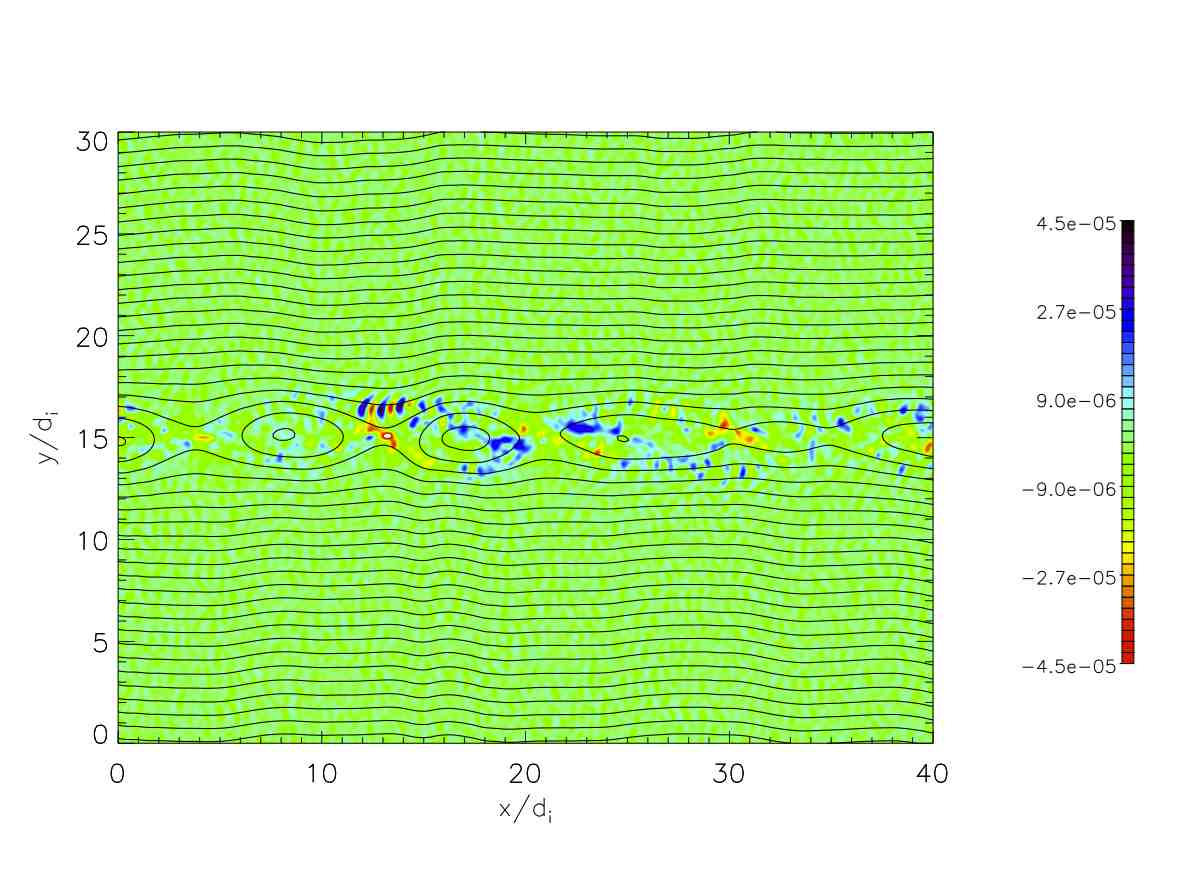}
\caption{Contour-plot of the electric field component parallel to the local magnetic field (parallel electric field  $E_{//} = (\mathbf{E} \cdot \mathbf{B})/| \mathbf{B} | $) on the $z = 0$ plane at time $\Omega_{ci} t = 4.8$. }
\label{Epar}
\end{figure}
The wave activity is shown in Fig.~\ref{Epar} as a wave-packet located at $11 < x/d_i < 15$ and $y/d_i \simeq 16$. In this figure, we show the parallel electric field $E_{//} = (\mathbf{E} \cdot \mathbf{B})/| \mathbf{B} | $ (the component of the electric field projected along the local magnetic field). 
When examining the electron fluid velocity in Fig.~\ref{VexGF}, the same region of space appears as a region of high velocity shear. However, an electron shearing instability has not been observed in this numerical experiment, at least until the end of the simulations.

In previous two dimensional simulations \citep{Markidis:2012}, the bipolar parallel electric field structures were found along the separatrices. In three dimension, the same signature does not appear to be located on the separatrices, at least in the plane of the two dimensional cut shown in Fig.~\ref{Epar}. This is due to the intrinsic three dimensional nature of the magnetic flux ropes, in particular the twisting of the plasmodia see Figs.~\ref{BlinesNOGF} and~\ref{B_GF}) is such that two nearby flux ropes do not face each other as two nearby magnetic islands do in two dimensional configuration~\citep{Markidis:2012}. Because of the more complex magnetic topology in three dimensions, the electron population accelerated by magnetic reconnection at two nearby $x$ lines do not necessarily encounter each other and therefore the electron two-stream instability is not expected to be triggered. On the contrary, the electron bump-on-tail or the Buneman instabilities are more likely to occur in three dimensional configurations with a guide field.

\begin{figure}[ht]
\includegraphics[width=0.95 \columnwidth]{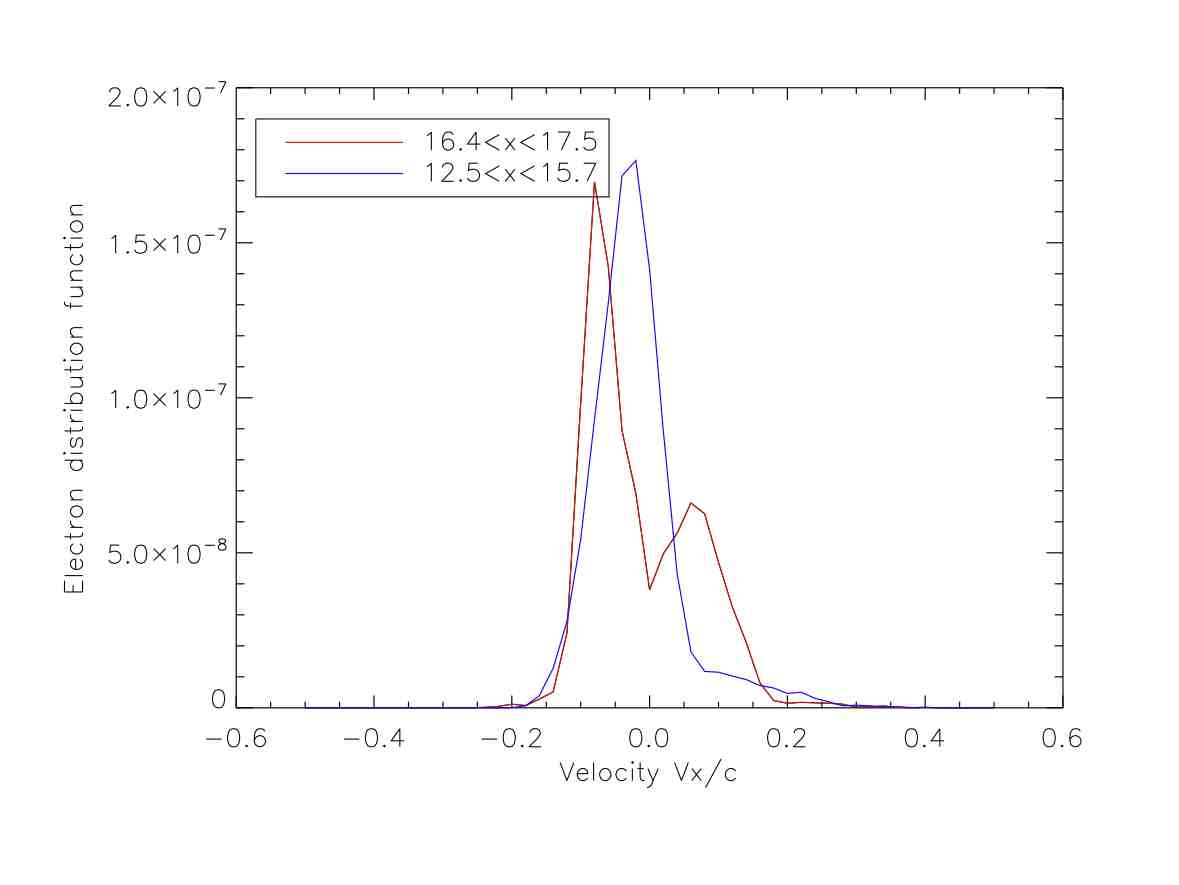}
\caption{Electron distribution functions $f_e(v_x)$ at time $\Omega_{ci} t = 4.8$ computed for $16.4 < x/d_i < 17.5$, $y/d_i \simeq 16.5$, $z/d_i \simeq 0.5$ (red line) and $12.5 < x/d_i < 15.7$, $y/d_i \simeq 16.5$, $z/d_i \simeq 0.5$ (blue line).
The electron beam visible in the electron distribution function in the region $16.4 < x/d_i < 17.5$ at $v_x/c \simeq 0.1$, relaxes to a plateau in the region $16.4 < x/d_i < 17.5$.}
\label{ElectronDistributionFunction}
\end{figure}

\begin{figure}[ht]
\includegraphics[width=0.95 \columnwidth]{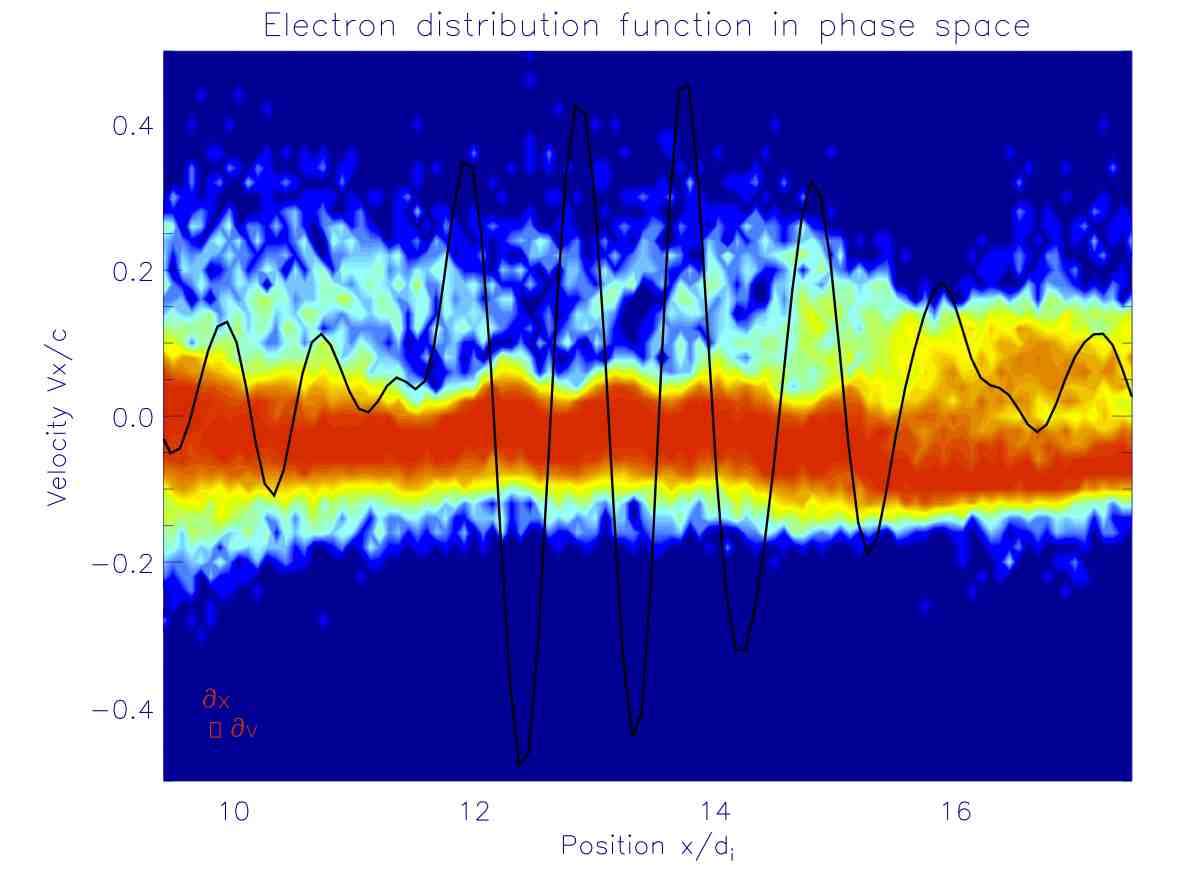}
\caption{Phase space $v_x/c - x/d_i$ at time $\Omega_{ci} t = 4.8$ in the region of space $16.4 < x/d_i < 17.5, < y < ,z=L_z/2$ with superimposed parallel electric field profile along the line in black color. The color table ranges from a low electron distribution function (blu for no electrons) to high distribution function (red). The binning used in the velocity and spatial directions are indicated with a red box in the bottom left corner. }
\label{PhaseSpace}
\end{figure}

To identify the nature of the kinetic instability at play, the electron distribution function $f_e(v_x)$ is computed at time $\Omega_{ci} t = 4.8$ and plotted in Fig.~\ref{ElectronDistributionFunction}, at two different locations: (i) at the edge of the wave-packet observed in Fig.~\ref{Epar} $12.5 < x/d_i < 15.7$ (blue line), (ii) at the center of the same wave-packet $16.4 < x/d_i < 17.5$ (red line). The presence of an electron beam is clearly identified in the electron distribution functions computed at the edge of the wave-packet (red line) at $v_x/c \simeq 0.1$, while the quasi-linear kinetic relaxation of the unstable distribution function is observed with the formation of a plateau in the (space-integrated) velocity space (blue line) at the location of the center of the wave-packet. This enables to identify the formation process of the wave-packet observed in Fig.~\ref{Epar} as an electron bump-on-tail instability.
To better understand the kinetic dynamics at play in this numerical experiment, the electron distribution function is computed in the phase space $v_x/c - x/d_i$ at time $\Omega_{ci} t = 4.8$, considering all the weighted electron computational particles in the cells surrounding the region of space $9.5 < x/d_i < 17.5$, $y/d_i \simeq 16.5$, $z/d_i \simeq 0.5$ and by integrating over velocity directions $v_y$ and $v_z$. The resulting electron distribution function is shown in Fig.~\ref{PhaseSpace}.
The two counter-streaming electron populations (electron core and beam) are visible in the region $16.4 < x/d_i < 17.5$, corresponding to the edge of the parallel electric field wave-packet, shown in Fig.~\ref{Epar} and over-plotted in black line. The presence of electron holes at positions $12 < x/d_i < 15$, $v_x/c \simeq 0.1$ corresponds to the electric potential wells of the wave-packet. The trapped electrons is a signature of the saturation of the bump-on-tail instability. It is seen as electron holes in the phase space (Fig.~\ref{PhaseSpace}) and as a plateau in the space-integrated velocity distribution function (Fig.~\ref{ElectronDistributionFunction}). We note that the electron holes form as a consequence of the bump-on-tail instability that develops at the later stages of our simulations. Streaming instabilities are not present at the early stages of plasmoid chain dynamics since initially the electron beams along the separatrices are not enough intense to trigger streaming instabilities~\citep{Markidis:2012}. The characteristics of electron holes along reconnection separatices have been discussed in previous studies reporting two and three dimensional Particle-in-Cell simulations~\citep{Lapenta:2010,LapentaGRL:2011,MarkidisPOP:2011}. It has been found that the typical size of the electron holes is approximately $16$ Debye lengths, and the bipolar structures are generated by streaming instability along reconnection separatrices. The presence of guide field and the use of different simulation charge to mass ratio influence the streaming instability along reconnection separatrices and the following formation of electron holes~\citep{Lapenta:2010}.

\section{Discussion}
The Particle-in-Cell simulations of plasmoid chain dynamics in three dimensions have been presented. Two Particle-in-Cell simulations of plasmoid chain evolution with and without a guide field allowed us to assess the effect of guide field in the evolution of the plasmoid chain. The overall dynamics for the two cases is similar in few aspects. In both cases, the tearing instability produces multiple $x$ lines magnetic reconnection sites. However, in the case of antiparallel magnetic reconnection a main reconnection site, comprising smaller ones, forms leading to an hierarchical structure of the current sheet. A hierarchical self-similar structure of the plasmoid-dominated current sheet was observed in previous fluid simulations~\citep{Shibata:2001, Uzdensky:2010}. This hierarchical structure in antiparallel reconnection might be caused by lower hybrid drift instability that leads to a non uniform thinning of the current sheet~\citep{Lapenta:2003}. In fact, waves in the range of the lower-hybrid frequency have been observed propagating along the $z$ axis in the simulations: early in time in the antiparallel configuration, and at the end of the simulation in presence of a guide field. 

It has been shown that a core field arises only in presence of a guide field, while it is absent when starting from an antiparallel set-up. This is in agreement with the idea that plasmoid core field originates from a compression of a pre-existing seed guide field~\citep{Karimabadi:1999, Drake:2006}. In addition, the core field of several plasmoids develops in the parallel to the initial guide field, suggesting the hypothesis that all the plasmoids originating from the same current sheet with uniform guide field present all the same polarity.

The presence of electron holes in plasmoid chain has been first revealed by two dimensional Particle-in-Cell simulations in Ref.~\citep{Markidis:2012} and it is now confirmed by three dimensional simulations. However, in the two dimensional geometry electron holes are localized along the separatrices and driven by two stream instabilities, while in the three dimensional case they are not necessarily along the separatrices and they are caused by bump-on-tail instability. We have discussed the occurrence of an electron kinetic instability, as a secondary instability of the kinetic tearing instability in presence of a guide magnetic field. When an initial guide magnetic field is not present, we do not observe similar bipolar parallel electric field structures in our simulation. Because of the huge computational resources needed in three dimensional kinetic simulations, we decided not to run the simulations for too long. It is thus not straightforward to conclude that the bipolar structures and streaming instability are absent in the case of antiparallel reconnection. Further numerical experiments running for longer simulation time will be necessary to properly evaluate the efficiency of kinetic electron streaming instabilities in three-dimensional geometry without a guide field.

In our simulations the central region of the magnetic islands is characterized by high density. In fact, plasmoids collect the outflow plasma from adjacent reconnection $x$ lines, progressively increasing their density. This is in agreement with some observations of density enhancement in the magnetic island~\citep{Slavin:2003,Retino:2008}, but it is not with recent studies where the density dips are observed in the central region of the plasmoid~\citep{Khotyaintsev:2010,Wang:2010}.  One possible explanation to this discrepancy is that the spacecrafts crossed regions of space where the coalescence phenomenon is ongoing. It has been shown in the present work and previous two dimensional simulations~\citep{Markidis:2012} that the density structures are rather complex in coalescing regions and they are characterized by relatively low density and electrostatic instabilities. Another possible explanation of the discrepancy between the simulation results and observations is the use of periodic boundary conditions in the $z$ direction that does not allow plasma to escape the plasmoids. Future simulations with open boundary conditions along the $z$ direction will clarify if the decrease of plasma density in the plasmoid is enabled by different boundary conditions.											\section{Conclusions}
The dynamics of a plasmoid chain in the kinetic framework have been studied by analyzing three dimensional Particle-in-Cell simulation results. It has been 
shown that the presence of a guide field dramatically influences the dynamics of the plasmoid chain. The plasmoid-dominated current sheet presents an hierarchical structure, where smaller reconnection sites are included in larger ones, in absence of a guide field, while plasmoids are approximately same size, and there is not a main reconnection site in the guide field simulations. A core field, parallel to the guide field, originates only in the case of guide field. The electron bump-on-tail instability is present in proximity of the plasmoids but not necessarily along separatrices, and bipolar structures of parallel electric field and electron holes are present in guide field reconnection. These results are relevant to study of multiple $x$ lines magnetic reconnection in Earth's magnetotail, revealing the three dimensional effects and pointing out that the presence of a moderate guide field modifies the signatures of plasmoid chain and multiple $x$ lines magnetic reconnection.

\begin{acknowledgments}
The present work is supported by NASA MMS Grant NNX08AO84G. Additional support for the KULeuven team is provided by the Onderzoekfonds KU Leuven (Research Fund KU Leuven) and by the European CommissionÕs Seventh Framework Programme (FP7/2007-2013) under the grant agreement no.~263340 (SWIFF project, www.swiff.eu). The KTH team acknowledges support from the European CommissionÕs Seventh Framework Programme under the grant agreement no.~287703 (CRESTA, cresta-project.eu). 
\end{acknowledgments}


\begin{thebibliography}{53}%
\makeatletter
\providecommand \@ifxundefined [1]{%
 \@ifx{#1\undefined}
}%
\providecommand \@ifnum [1]{%
 \ifnum #1\expandafter \@firstoftwo
 \else \expandafter \@secondoftwo
 \fi
}%
\providecommand \@ifx [1]{%
 \ifx #1\expandafter \@firstoftwo
 \else \expandafter \@secondoftwo
 \fi
}%
\providecommand \natexlab [1]{#1}%
\providecommand \enquote  [1]{``#1''}%
\providecommand \bibnamefont  [1]{#1}%
\providecommand \bibfnamefont [1]{#1}%
\providecommand \citenamefont [1]{#1}%
\providecommand \href@noop [0]{\@secondoftwo}%
\providecommand \href [0]{\begingroup \@sanitize@url \@href}%
\providecommand \@href[1]{\@@startlink{#1}\@@href}%
\providecommand \@@href[1]{\endgroup#1\@@endlink}%
\providecommand \@sanitize@url [0]{\catcode `\\12\catcode `\$12\catcode
  `\&12\catcode `\#12\catcode `\^12\catcode `\_12\catcode `\%12\relax}%
\providecommand \@@startlink[1]{}%
\providecommand \@@endlink[0]{}%
\providecommand \url  [0]{\begingroup\@sanitize@url \@url }%
\providecommand \@url [1]{\endgroup\@href {#1}{\urlprefix }}%
\providecommand \urlprefix  [0]{URL }%
\providecommand \Eprint [0]{\href }%
\providecommand \doibase [0]{http://dx.doi.org/}%
\providecommand \selectlanguage [0]{\@gobble}%
\providecommand \bibinfo  [0]{\@secondoftwo}%
\providecommand \bibfield  [0]{\@secondoftwo}%
\providecommand \translation [1]{[#1]}%
\providecommand \BibitemOpen [0]{}%
\providecommand \bibitemStop [0]{}%
\providecommand \bibitemNoStop [0]{.\EOS\space}%
\providecommand \EOS [0]{\spacefactor3000\relax}%
\providecommand \BibitemShut  [1]{\csname bibitem#1\endcsname}%
\let\auto@bib@innerbib\@empty
\bibitem [{\citenamefont {Hughes}\ and\ \citenamefont
  {Sibeck}(1987)}]{Hughes:1987}%
  \BibitemOpen
  \bibfield  {author} {\bibinfo {author} {\bibfnamefont {W.}~\bibnamefont
  {Hughes}}\ and\ \bibinfo {author} {\bibfnamefont {D.}~\bibnamefont
  {Sibeck}},\ }\href@noop {} {\bibfield  {journal} {\bibinfo  {journal}
  {Geophysical Research Letters}\ }\textbf {\bibinfo {volume} {14}},\ \bibinfo
  {pages} {636} (\bibinfo {year} {1987})}\BibitemShut {NoStop}%
\bibitem [{\citenamefont {Slavin}\ \emph {et~al.}(2003)\citenamefont {Slavin},
  \citenamefont {Lepping}, \citenamefont {Gjerloev}, \citenamefont {Fairfield},
  \citenamefont {Hesse}, \citenamefont {Owen}, \citenamefont {Moldwin},
  \citenamefont {Nagai}, \citenamefont {Ieda},\ and\ \citenamefont
  {Mukai}}]{Slavin:2003}%
  \BibitemOpen
  \bibfield  {author} {\bibinfo {author} {\bibfnamefont {J.}~\bibnamefont
  {Slavin}}, \bibinfo {author} {\bibfnamefont {R.}~\bibnamefont {Lepping}},
  \bibinfo {author} {\bibfnamefont {J.}~\bibnamefont {Gjerloev}}, \bibinfo
  {author} {\bibfnamefont {D.}~\bibnamefont {Fairfield}}, \bibinfo {author}
  {\bibfnamefont {M.}~\bibnamefont {Hesse}}, \bibinfo {author} {\bibfnamefont
  {C.}~\bibnamefont {Owen}}, \bibinfo {author} {\bibfnamefont {M.}~\bibnamefont
  {Moldwin}}, \bibinfo {author} {\bibfnamefont {T.}~\bibnamefont {Nagai}},
  \bibinfo {author} {\bibfnamefont {A.}~\bibnamefont {Ieda}}, \ and\ \bibinfo
  {author} {\bibfnamefont {T.}~\bibnamefont {Mukai}},\ }\href@noop {}
  {\bibfield  {journal} {\bibinfo  {journal} {Journal of Geophysical Research}\
  }\textbf {\bibinfo {volume} {108}},\ \bibinfo {pages} {1015} (\bibinfo {year}
  {2003})}\BibitemShut {NoStop}%
\bibitem [{\citenamefont {{Hones}}(1977)}]{Hones:1977}%
  \BibitemOpen
  \bibfield  {author} {\bibinfo {author} {\bibfnamefont {E.~W.}\ \bibnamefont
  {{Hones}}, \bibfnamefont {Jr.}},\ }\href@noop {} {\bibfield  {journal}
  {\bibinfo  {journal} {Journal of Geophysical Research (Space Physics)}\
  }\textbf {\bibinfo {volume} {82}},\ \bibinfo {pages} {5633} (\bibinfo {year}
  {1977})}\BibitemShut {NoStop}%
\bibitem [{\citenamefont {Bhattacharjee}\ \emph {et~al.}(1983)\citenamefont
  {Bhattacharjee}, \citenamefont {Brunel},\ and\ \citenamefont
  {Tajima}}]{Bhattacharjee:1983}%
  \BibitemOpen
  \bibfield  {author} {\bibinfo {author} {\bibfnamefont {A.}~\bibnamefont
  {Bhattacharjee}}, \bibinfo {author} {\bibfnamefont {F.}~\bibnamefont
  {Brunel}}, \ and\ \bibinfo {author} {\bibfnamefont {T.}~\bibnamefont
  {Tajima}},\ }\href@noop {} {\bibfield  {journal} {\bibinfo  {journal}
  {Physics of Fluids}\ }\textbf {\bibinfo {volume} {26}},\ \bibinfo {pages}
  {3332} (\bibinfo {year} {1983})}\BibitemShut {NoStop}%
\bibitem [{\citenamefont {{Intrator}}\ \emph {et~al.}(2009)\citenamefont
  {{Intrator}}, \citenamefont {{Sun}}, \citenamefont {{Lapenta}}, \citenamefont
  {{Dorf}},\ and\ \citenamefont {{Furno}}}]{Intrator:2009}%
  \BibitemOpen
  \bibfield  {author} {\bibinfo {author} {\bibfnamefont {T.~P.}\ \bibnamefont
  {{Intrator}}}, \bibinfo {author} {\bibfnamefont {X.}~\bibnamefont {{Sun}}},
  \bibinfo {author} {\bibfnamefont {G.}~\bibnamefont {{Lapenta}}}, \bibinfo
  {author} {\bibfnamefont {L.}~\bibnamefont {{Dorf}}}, \ and\ \bibinfo {author}
  {\bibfnamefont {I.}~\bibnamefont {{Furno}}},\ }\href@noop {} {\bibfield
  {journal} {\bibinfo  {journal} {Nature Physics}\ }\textbf {\bibinfo {volume}
  {5}},\ \bibinfo {pages} {521} (\bibinfo {year} {2009})}\BibitemShut {NoStop}%
\bibitem [{\citenamefont {Karimabadi}\ \emph {et~al.}(2011)\citenamefont
  {Karimabadi}, \citenamefont {Dorelli}, \citenamefont {Roytershteyn},
  \citenamefont {Daughton},\ and\ \citenamefont {Chac\'on}}]{Karimabadi:2011}%
  \BibitemOpen
  \bibfield  {author} {\bibinfo {author} {\bibfnamefont {H.}~\bibnamefont
  {Karimabadi}}, \bibinfo {author} {\bibfnamefont {J.}~\bibnamefont {Dorelli}},
  \bibinfo {author} {\bibfnamefont {V.}~\bibnamefont {Roytershteyn}}, \bibinfo
  {author} {\bibfnamefont {W.}~\bibnamefont {Daughton}}, \ and\ \bibinfo
  {author} {\bibfnamefont {L.}~\bibnamefont {Chac\'on}},\ }\href@noop {}
  {\bibfield  {journal} {\bibinfo  {journal} {Phys. Rev. Lett.}\ }\textbf
  {\bibinfo {volume} {107}},\ \bibinfo {pages} {025002} (\bibinfo {year}
  {2011})}\BibitemShut {NoStop}%
\bibitem [{\citenamefont {Daughton}\ \emph {et~al.}(2009)\citenamefont
  {Daughton}, \citenamefont {Roytershteyn}, \citenamefont {Albright},
  \citenamefont {Karimabadi}, \citenamefont {Yin},\ and\ \citenamefont
  {Bowers}}]{Daughton:2009}%
  \BibitemOpen
  \bibfield  {author} {\bibinfo {author} {\bibfnamefont {W.}~\bibnamefont
  {Daughton}}, \bibinfo {author} {\bibfnamefont {V.}~\bibnamefont
  {Roytershteyn}}, \bibinfo {author} {\bibfnamefont {B.~J.}\ \bibnamefont
  {Albright}}, \bibinfo {author} {\bibfnamefont {H.}~\bibnamefont
  {Karimabadi}}, \bibinfo {author} {\bibfnamefont {L.}~\bibnamefont {Yin}}, \
  and\ \bibinfo {author} {\bibfnamefont {K.~J.}\ \bibnamefont {Bowers}},\
  }\href {\doibase 10.1103/PhysRevLett.103.065004} {\bibfield  {journal}
  {\bibinfo  {journal} {Phys. Rev. Lett.}\ }\textbf {\bibinfo {volume} {103}},\
  \bibinfo {pages} {065004} (\bibinfo {year} {2009})}\BibitemShut {NoStop}%
\bibitem [{\citenamefont {{Lin}}\ \emph {et~al.}(2008)\citenamefont {{Lin}},
  \citenamefont {{Cranmer}},\ and\ \citenamefont {{Farrugia}}}]{Lin:2008}%
  \BibitemOpen
  \bibfield  {author} {\bibinfo {author} {\bibfnamefont {J.}~\bibnamefont
  {{Lin}}}, \bibinfo {author} {\bibfnamefont {S.~R.}\ \bibnamefont
  {{Cranmer}}}, \ and\ \bibinfo {author} {\bibfnamefont {C.~J.}\ \bibnamefont
  {{Farrugia}}},\ }\href@noop {} {\bibfield  {journal} {\bibinfo  {journal}
  {Journal of Geophysical Research (Space Physics)}\ }\textbf {\bibinfo
  {volume} {113}},\ \bibinfo {pages} {A11107} (\bibinfo {year}
  {2008})}\BibitemShut {NoStop}%
\bibitem [{\citenamefont {Khotyaintsev}\ \emph {et~al.}(2010)\citenamefont
  {Khotyaintsev}, \citenamefont {Vaivads}, \citenamefont {Andr\'e},
  \citenamefont {Fujimoto}, \citenamefont {Retin\`o},\ and\ \citenamefont
  {Owen}}]{Khotyaintsev:2010}%
  \BibitemOpen
  \bibfield  {author} {\bibinfo {author} {\bibfnamefont {Y.~V.}\ \bibnamefont
  {Khotyaintsev}}, \bibinfo {author} {\bibfnamefont {A.}~\bibnamefont
  {Vaivads}}, \bibinfo {author} {\bibfnamefont {M.}~\bibnamefont {Andr\'e}},
  \bibinfo {author} {\bibfnamefont {M.}~\bibnamefont {Fujimoto}}, \bibinfo
  {author} {\bibfnamefont {A.}~\bibnamefont {Retin\`o}}, \ and\ \bibinfo
  {author} {\bibfnamefont {C.~J.}\ \bibnamefont {Owen}},\ }\href@noop {}
  {\bibfield  {journal} {\bibinfo  {journal} {Phys. Rev. Lett.}\ }\textbf
  {\bibinfo {volume} {105}},\ \bibinfo {pages} {165002} (\bibinfo {year}
  {2010})}\BibitemShut {NoStop}%
\bibitem [{\citenamefont {Borg}\ \emph {et~al.}(2012)\citenamefont {Borg},
  \citenamefont {Taylor},\ and\ \citenamefont {Eastwood}}]{Borg:2012}%
  \BibitemOpen
  \bibfield  {author} {\bibinfo {author} {\bibfnamefont {A.}~\bibnamefont
  {Borg}}, \bibinfo {author} {\bibfnamefont {M.}~\bibnamefont {Taylor}}, \ and\
  \bibinfo {author} {\bibfnamefont {J.}~\bibnamefont {Eastwood}},\ }\href@noop
  {} {\bibfield  {journal} {\bibinfo  {journal} {Annales
  Geophysicae-Atmospheres Hydrospheresand Space Sciences}\ }\textbf {\bibinfo
  {volume} {30}},\ \bibinfo {pages} {761} (\bibinfo {year} {2012})}\BibitemShut
  {NoStop}%
\bibitem [{\citenamefont {Slavin}\ \emph {et~al.}(2012)\citenamefont {Slavin},
  \citenamefont {Imber}, \citenamefont {Boardsen}, \citenamefont {DiBraccio},
  \citenamefont {Sundberg}, \citenamefont {Sarantos}, \citenamefont
  {Nieves-Chinchilla}, \citenamefont {Szabo}, \citenamefont {Anderson},
  \citenamefont {Korth} \emph {et~al.}}]{slavin2012messenger}%
  \BibitemOpen
  \bibfield  {author} {\bibinfo {author} {\bibfnamefont {J.~A.}\ \bibnamefont
  {Slavin}}, \bibinfo {author} {\bibfnamefont {S.~M.}\ \bibnamefont {Imber}},
  \bibinfo {author} {\bibfnamefont {S.~A.}\ \bibnamefont {Boardsen}}, \bibinfo
  {author} {\bibfnamefont {G.~A.}\ \bibnamefont {DiBraccio}}, \bibinfo {author}
  {\bibfnamefont {T.}~\bibnamefont {Sundberg}}, \bibinfo {author}
  {\bibfnamefont {M.}~\bibnamefont {Sarantos}}, \bibinfo {author}
  {\bibfnamefont {T.}~\bibnamefont {Nieves-Chinchilla}}, \bibinfo {author}
  {\bibfnamefont {A.}~\bibnamefont {Szabo}}, \bibinfo {author} {\bibfnamefont
  {B.~J.}\ \bibnamefont {Anderson}}, \bibinfo {author} {\bibfnamefont
  {H.}~\bibnamefont {Korth}},  \emph {et~al.},\ }\href@noop {} {\bibfield
  {journal} {\bibinfo  {journal} {Journal of Geophysical Research: Space
  Physics (1978--2012)}\ }\textbf {\bibinfo {volume} {117}} (\bibinfo {year}
  {2012})}\BibitemShut {NoStop}%
\bibitem [{\citenamefont {{Bemporad}}(2008)}]{Bemporad:2008}%
  \BibitemOpen
  \bibfield  {author} {\bibinfo {author} {\bibfnamefont {A.}~\bibnamefont
  {{Bemporad}}},\ }\href@noop {} {\bibfield  {journal} {\bibinfo  {journal}
  {Astrophysical Journal}\ }\textbf {\bibinfo {volume} {689}},\ \bibinfo
  {pages} {572} (\bibinfo {year} {2008})}\BibitemShut {NoStop}%
\bibitem [{\citenamefont {Henri}\ \emph {et~al.}(2012)\citenamefont {Henri},
  \citenamefont {Califano}, \citenamefont {Faganello},\ and\ \citenamefont
  {Pegoraro}}]{Henri:2012}%
  \BibitemOpen
  \bibfield  {author} {\bibinfo {author} {\bibfnamefont {P.}~\bibnamefont
  {Henri}}, \bibinfo {author} {\bibfnamefont {F.}~\bibnamefont {Califano}},
  \bibinfo {author} {\bibfnamefont {M.}~\bibnamefont {Faganello}}, \ and\
  \bibinfo {author} {\bibfnamefont {F.}~\bibnamefont {Pegoraro}},\ }\href@noop
  {} {\bibfield  {journal} {\bibinfo  {journal} {Physics of Plasmas}\ }\textbf
  {\bibinfo {volume} {19}},\ \bibinfo {pages} {072908} (\bibinfo {year}
  {2012})}\BibitemShut {NoStop}%
\bibitem [{\citenamefont {{Biskamp}}(1986)}]{Biskamp:1986}%
  \BibitemOpen
  \bibfield  {author} {\bibinfo {author} {\bibfnamefont {D.}~\bibnamefont
  {{Biskamp}}},\ }\href@noop {} {\bibfield  {journal} {\bibinfo  {journal}
  {Phys. Fluids}\ }\textbf {\bibinfo {volume} {29}},\ \bibinfo {pages} {1520}
  (\bibinfo {year} {1986})}\BibitemShut {NoStop}%
\bibitem [{\citenamefont {Malara}\ \emph {et~al.}(1992)\citenamefont {Malara},
  \citenamefont {Veltri},\ and\ \citenamefont {Carbone}}]{Malara:1992}%
  \BibitemOpen
  \bibfield  {author} {\bibinfo {author} {\bibfnamefont {F.}~\bibnamefont
  {Malara}}, \bibinfo {author} {\bibfnamefont {P.}~\bibnamefont {Veltri}}, \
  and\ \bibinfo {author} {\bibfnamefont {V.}~\bibnamefont {Carbone}},\
  }\href@noop {} {\bibfield  {journal} {\bibinfo  {journal} {Physics of
  Plasmas}\ }\textbf {\bibinfo {volume} {4}},\ \bibinfo {pages} {3070}
  (\bibinfo {year} {1992})}\BibitemShut {NoStop}%
\bibitem [{\citenamefont {{Lazarian}}\ and\ \citenamefont
  {{Vishniac}}(1999)}]{Lazarian:1999}%
  \BibitemOpen
  \bibfield  {author} {\bibinfo {author} {\bibfnamefont {A.}~\bibnamefont
  {{Lazarian}}}\ and\ \bibinfo {author} {\bibfnamefont {E.~T.}\ \bibnamefont
  {{Vishniac}}},\ }\href@noop {} {\bibfield  {journal} {\bibinfo  {journal}
  {Astrophysical Journal}\ }\textbf {\bibinfo {volume} {517}},\ \bibinfo
  {pages} {700} (\bibinfo {year} {1999})}\BibitemShut {NoStop}%
\bibitem [{\citenamefont {{Kowal}}\ \emph {et~al.}(2009)\citenamefont
  {{Kowal}}, \citenamefont {{Lazarian}}, \citenamefont {{Vishniac}},\ and\
  \citenamefont {{Otmianowska-Mazur}}}]{Kowal:2009}%
  \BibitemOpen
  \bibfield  {author} {\bibinfo {author} {\bibfnamefont {G.}~\bibnamefont
  {{Kowal}}}, \bibinfo {author} {\bibfnamefont {A.}~\bibnamefont {{Lazarian}}},
  \bibinfo {author} {\bibfnamefont {E.~T.}\ \bibnamefont {{Vishniac}}}, \ and\
  \bibinfo {author} {\bibfnamefont {K.}~\bibnamefont {{Otmianowska-Mazur}}},\
  }\href@noop {} {\bibfield  {journal} {\bibinfo  {journal} {Astrophysical
  Journal}\ }\textbf {\bibinfo {volume} {700}},\ \bibinfo {pages} {63}
  (\bibinfo {year} {2009})}\BibitemShut {NoStop}%
\bibitem [{\citenamefont {Loureiro}\ \emph {et~al.}(2005)\citenamefont
  {Loureiro}, \citenamefont {Cowley}, \citenamefont {Dorland}, \citenamefont
  {Haines},\ and\ \citenamefont {Schekochihin}}]{Loureiro:2005}%
  \BibitemOpen
  \bibfield  {author} {\bibinfo {author} {\bibfnamefont {N.~F.}\ \bibnamefont
  {Loureiro}}, \bibinfo {author} {\bibfnamefont {S.~C.}\ \bibnamefont
  {Cowley}}, \bibinfo {author} {\bibfnamefont {W.~D.}\ \bibnamefont {Dorland}},
  \bibinfo {author} {\bibfnamefont {M.~G.}\ \bibnamefont {Haines}}, \ and\
  \bibinfo {author} {\bibfnamefont {A.~A.}\ \bibnamefont {Schekochihin}},\
  }\href {\doibase 10.1103/PhysRevLett.95.235003} {\bibfield  {journal}
  {\bibinfo  {journal} {Phys. Rev. Lett.}\ }\textbf {\bibinfo {volume} {95}},\
  \bibinfo {pages} {235003} (\bibinfo {year} {2005})}\BibitemShut {NoStop}%
\bibitem [{\citenamefont {Loureiro}\ \emph {et~al.}(2007)\citenamefont
  {Loureiro}, \citenamefont {Schekochihin},\ and\ \citenamefont
  {Cowley}}]{Loureiro:2007}%
  \BibitemOpen
  \bibfield  {author} {\bibinfo {author} {\bibfnamefont {N.~F.}\ \bibnamefont
  {Loureiro}}, \bibinfo {author} {\bibfnamefont {A.~A.}\ \bibnamefont
  {Schekochihin}}, \ and\ \bibinfo {author} {\bibfnamefont {S.~C.}\
  \bibnamefont {Cowley}},\ }\href@noop {} {\bibfield  {journal} {\bibinfo
  {journal} {Physics of Plasmas}\ }\textbf {\bibinfo {volume} {14}},\ \bibinfo
  {eid} {100703} (\bibinfo {year} {2007})}\BibitemShut {NoStop}%
\bibitem [{\citenamefont {Lapenta}(2008)}]{Lapenta:2008}%
  \BibitemOpen
  \bibfield  {author} {\bibinfo {author} {\bibfnamefont {G.}~\bibnamefont
  {Lapenta}},\ }\href {\doibase 10.1103/PhysRevLett.100.235001} {\bibfield
  {journal} {\bibinfo  {journal} {Phys. Rev. Lett.}\ }\textbf {\bibinfo
  {volume} {100}},\ \bibinfo {pages} {235001} (\bibinfo {year}
  {2008})}\BibitemShut {NoStop}%
\bibitem [{\citenamefont {Bhattacharjee}\ \emph {et~al.}(2009)\citenamefont
  {Bhattacharjee}, \citenamefont {Huang}, \citenamefont {Yang},\ and\
  \citenamefont {Rogers}}]{bhattacharjee:2009}%
  \BibitemOpen
  \bibfield  {author} {\bibinfo {author} {\bibfnamefont {A.}~\bibnamefont
  {Bhattacharjee}}, \bibinfo {author} {\bibfnamefont {Y.-M.}\ \bibnamefont
  {Huang}}, \bibinfo {author} {\bibfnamefont {H.}~\bibnamefont {Yang}}, \ and\
  \bibinfo {author} {\bibfnamefont {B.}~\bibnamefont {Rogers}},\ }\href@noop {}
  {\bibfield  {journal} {\bibinfo  {journal} {Physics of Plasmas}\ }\textbf
  {\bibinfo {volume} {16}},\ \bibinfo {eid} {112102} (\bibinfo {year}
  {2009})}\BibitemShut {NoStop}%
\bibitem [{\citenamefont {Cassak}\ \emph {et~al.}(2009)\citenamefont {Cassak},
  \citenamefont {Shay},\ and\ \citenamefont {Drake}}]{Cassak:2009}%
  \BibitemOpen
  \bibfield  {author} {\bibinfo {author} {\bibfnamefont {P.~A.}\ \bibnamefont
  {Cassak}}, \bibinfo {author} {\bibfnamefont {M.~A.}\ \bibnamefont {Shay}}, \
  and\ \bibinfo {author} {\bibfnamefont {J.~F.}\ \bibnamefont {Drake}},\
  }\href@noop {} {\bibfield  {journal} {\bibinfo  {journal} {Physics of
  Plasmas}\ }\textbf {\bibinfo {volume} {16}},\ \bibinfo {pages} {120702}
  (\bibinfo {year} {2009})}\BibitemShut {NoStop}%
\bibitem [{\citenamefont {Uzdensky}\ \emph {et~al.}(2010)\citenamefont
  {Uzdensky}, \citenamefont {Loureiro},\ and\ \citenamefont
  {Schekochihin}}]{Uzdensky:2010}%
  \BibitemOpen
  \bibfield  {author} {\bibinfo {author} {\bibfnamefont {D.~A.}\ \bibnamefont
  {Uzdensky}}, \bibinfo {author} {\bibfnamefont {N.~F.}\ \bibnamefont
  {Loureiro}}, \ and\ \bibinfo {author} {\bibfnamefont {A.~A.}\ \bibnamefont
  {Schekochihin}},\ }\href@noop {} {\bibfield  {journal} {\bibinfo  {journal}
  {Phys. Rev. Lett.}\ }\textbf {\bibinfo {volume} {105}},\ \bibinfo {pages}
  {235002} (\bibinfo {year} {2010})}\BibitemShut {NoStop}%
\bibitem [{\citenamefont {{Tajima}}\ \emph {et~al.}(1987)\citenamefont
  {{Tajima}}, \citenamefont {{Sakai}}, \citenamefont {{Nakajima}},
  \citenamefont {{Kosugi}}, \citenamefont {{Brunel}},\ and\ \citenamefont
  {{Kundu}}}]{Tajima:1987}%
  \BibitemOpen
  \bibfield  {author} {\bibinfo {author} {\bibfnamefont {T.}~\bibnamefont
  {{Tajima}}}, \bibinfo {author} {\bibfnamefont {J.}~\bibnamefont {{Sakai}}},
  \bibinfo {author} {\bibfnamefont {H.}~\bibnamefont {{Nakajima}}}, \bibinfo
  {author} {\bibfnamefont {T.}~\bibnamefont {{Kosugi}}}, \bibinfo {author}
  {\bibfnamefont {F.}~\bibnamefont {{Brunel}}}, \ and\ \bibinfo {author}
  {\bibfnamefont {M.~R.}\ \bibnamefont {{Kundu}}},\ }\href@noop {} {\bibfield
  {journal} {\bibinfo  {journal} {Astrophysical Journal}\ }\textbf {\bibinfo
  {volume} {321}},\ \bibinfo {pages} {1031} (\bibinfo {year}
  {1987})}\BibitemShut {NoStop}%
\bibitem [{\citenamefont {Pritchett}(2008)}]{Pritchett:2008}%
  \BibitemOpen
  \bibfield  {author} {\bibinfo {author} {\bibfnamefont {P.~L.}\ \bibnamefont
  {Pritchett}},\ }\href@noop {} {\bibfield  {journal} {\bibinfo  {journal}
  {Physics of Plasmas}\ }\textbf {\bibinfo {volume} {15}},\ \bibinfo {pages}
  {102105} (\bibinfo {year} {2008})}\BibitemShut {NoStop}%
\bibitem [{\citenamefont {{Oka}}\ \emph
  {et~al.}(2010{\natexlab{a}})\citenamefont {{Oka}}, \citenamefont {{Phan}},
  \citenamefont {{Krucker}}, \citenamefont {{Fujimoto}},\ and\ \citenamefont
  {{Shinohara}}}]{OkaAPJ:2010}%
  \BibitemOpen
  \bibfield  {author} {\bibinfo {author} {\bibfnamefont {M.}~\bibnamefont
  {{Oka}}}, \bibinfo {author} {\bibfnamefont {T.-D.}\ \bibnamefont {{Phan}}},
  \bibinfo {author} {\bibfnamefont {S.}~\bibnamefont {{Krucker}}}, \bibinfo
  {author} {\bibfnamefont {M.}~\bibnamefont {{Fujimoto}}}, \ and\ \bibinfo
  {author} {\bibfnamefont {I.}~\bibnamefont {{Shinohara}}},\ }\href@noop {}
  {\bibfield  {journal} {\bibinfo  {journal} {Astrophysical Journal}\ }\textbf
  {\bibinfo {volume} {714}},\ \bibinfo {pages} {915} (\bibinfo {year}
  {2010}{\natexlab{a}})}\BibitemShut {NoStop}%
\bibitem [{\citenamefont {{Oka}}\ \emph
  {et~al.}(2010{\natexlab{b}})\citenamefont {{Oka}}, \citenamefont
  {{Fujimoto}}, \citenamefont {{Shinohara}},\ and\ \citenamefont
  {{Phan}}}]{OkaJGR:2010}%
  \BibitemOpen
  \bibfield  {author} {\bibinfo {author} {\bibfnamefont {M.}~\bibnamefont
  {{Oka}}}, \bibinfo {author} {\bibfnamefont {M.}~\bibnamefont {{Fujimoto}}},
  \bibinfo {author} {\bibfnamefont {I.}~\bibnamefont {{Shinohara}}}, \ and\
  \bibinfo {author} {\bibfnamefont {T.~D.}\ \bibnamefont {{Phan}}},\
  }\href@noop {} {\bibfield  {journal} {\bibinfo  {journal} {Journal of
  Geophysical Research (Space Physics)}\ }\textbf {\bibinfo {volume} {115}},\
  \bibinfo {pages} {A08223} (\bibinfo {year} {2010}{\natexlab{b}})}\BibitemShut
  {NoStop}%
\bibitem [{\citenamefont {Tanaka}\ \emph {et~al.}(2010)\citenamefont {Tanaka},
  \citenamefont {Yumura}, \citenamefont {Fujimoto}, \citenamefont {Shinohara},
  \citenamefont {Badman},\ and\ \citenamefont {Grocott}}]{Tanaka:2010}%
  \BibitemOpen
  \bibfield  {author} {\bibinfo {author} {\bibfnamefont {K.~G.}\ \bibnamefont
  {Tanaka}}, \bibinfo {author} {\bibfnamefont {T.}~\bibnamefont {Yumura}},
  \bibinfo {author} {\bibfnamefont {M.}~\bibnamefont {Fujimoto}}, \bibinfo
  {author} {\bibfnamefont {I.}~\bibnamefont {Shinohara}}, \bibinfo {author}
  {\bibfnamefont {S.~V.}\ \bibnamefont {Badman}}, \ and\ \bibinfo {author}
  {\bibfnamefont {A.}~\bibnamefont {Grocott}},\ }\href@noop {} {\bibfield
  {journal} {\bibinfo  {journal} {Physics of Plasmas}\ }\textbf {\bibinfo
  {volume} {17}},\ \bibinfo {pages} {102902} (\bibinfo {year}
  {2010})}\BibitemShut {NoStop}%
\bibitem [{\citenamefont {{Markidis}}\ \emph {et~al.}(2012)\citenamefont
  {{Markidis}}, \citenamefont {{Henri}}, \citenamefont {{Lapenta}},
  \citenamefont {{Divin}}, \citenamefont {{Goldman}}, \citenamefont
  {{Newman}},\ and\ \citenamefont {{Eriksson}}}]{Markidis:2012}%
  \BibitemOpen
  \bibfield  {author} {\bibinfo {author} {\bibfnamefont {S.}~\bibnamefont
  {{Markidis}}}, \bibinfo {author} {\bibfnamefont {P.}~\bibnamefont {{Henri}}},
  \bibinfo {author} {\bibfnamefont {G.}~\bibnamefont {{Lapenta}}}, \bibinfo
  {author} {\bibfnamefont {A.}~\bibnamefont {{Divin}}}, \bibinfo {author}
  {\bibfnamefont {M.~V.}\ \bibnamefont {{Goldman}}}, \bibinfo {author}
  {\bibfnamefont {D.}~\bibnamefont {{Newman}}}, \ and\ \bibinfo {author}
  {\bibfnamefont {S.}~\bibnamefont {{Eriksson}}},\ }\href {\doibase
  10.5194/npg-19-145-2012} {\bibfield  {journal} {\bibinfo  {journal}
  {Nonlinear Processes in Geophysics}\ }\textbf {\bibinfo {volume} {19}},\
  \bibinfo {pages} {145} (\bibinfo {year} {2012})}\BibitemShut {NoStop}%
\bibitem [{\citenamefont {{Hoshino}}(2012)}]{Hoshino:2012}%
  \BibitemOpen
  \bibfield  {author} {\bibinfo {author} {\bibfnamefont {M.}~\bibnamefont
  {{Hoshino}}},\ }\href {\doibase 10.1103/PhysRevLett.108.135003} {\bibfield
  {journal} {\bibinfo  {journal} {Physical Review Letters}\ }\textbf {\bibinfo
  {volume} {108}},\ \bibinfo {eid} {135003} (\bibinfo {year}
  {2012})}\BibitemShut {NoStop}%
\bibitem [{\citenamefont {{Drake}}\ \emph
  {et~al.}(2006{\natexlab{a}})\citenamefont {{Drake}}, \citenamefont
  {{Swisdak}}, \citenamefont {{Che}},\ and\ \citenamefont
  {{Shay}}}]{Drake:Nature}%
  \BibitemOpen
  \bibfield  {author} {\bibinfo {author} {\bibfnamefont {J.~F.}\ \bibnamefont
  {{Drake}}}, \bibinfo {author} {\bibfnamefont {M.}~\bibnamefont {{Swisdak}}},
  \bibinfo {author} {\bibfnamefont {H.}~\bibnamefont {{Che}}}, \ and\ \bibinfo
  {author} {\bibfnamefont {M.~A.}\ \bibnamefont {{Shay}}},\ }\href@noop {}
  {\bibfield  {journal} {\bibinfo  {journal} {Nature Physics}\ }\textbf
  {\bibinfo {volume} {443}},\ \bibinfo {pages} {553} (\bibinfo {year}
  {2006}{\natexlab{a}})}\BibitemShut {NoStop}%
\bibitem [{\citenamefont {{Kagan}}\ \emph {et~al.}(2012)\citenamefont
  {{Kagan}}, \citenamefont {{Milosavljevic}},\ and\ \citenamefont
  {{Spitkovsky}}}]{Kagan:2012}%
  \BibitemOpen
  \bibfield  {author} {\bibinfo {author} {\bibfnamefont {D.}~\bibnamefont
  {{Kagan}}}, \bibinfo {author} {\bibfnamefont {M.}~\bibnamefont
  {{Milosavljevic}}}, \ and\ \bibinfo {author} {\bibfnamefont {A.}~\bibnamefont
  {{Spitkovsky}}},\ }\href@noop {} {\bibfield  {journal} {\bibinfo  {journal}
  {ArXiv e-prints}\ } (\bibinfo {year} {2012})},\ \Eprint
  {http://arxiv.org/abs/1208.0849} {1208.0849 [astro-ph.HE]} \BibitemShut
  {NoStop}%
\bibitem [{\citenamefont {{Wang}}\ \emph {et~al.}(2012)\citenamefont {{Wang}},
  \citenamefont {{Nakamura}}, \citenamefont {{Lu}}, \citenamefont {{Du}},
  \citenamefont {{Zhang}}, \citenamefont {{Baumjohann}}, \citenamefont
  {{Khotyaintsev}}, \citenamefont {{Volwerk}}, \citenamefont {{Andr{\'e}}},
  \citenamefont {{Fujimoto}}, \citenamefont {{Nakamura}}, \citenamefont
  {{Fazakerley}}, \citenamefont {{Du}}, \citenamefont {{Teh}}, \citenamefont
  {{Panov}}, \citenamefont {{Zieger}}, \citenamefont {{Pan}},\ and\
  \citenamefont {{Lu}}}]{Wang:2012}%
  \BibitemOpen
  \bibfield  {author} {\bibinfo {author} {\bibfnamefont {R.}~\bibnamefont
  {{Wang}}}, \bibinfo {author} {\bibfnamefont {R.}~\bibnamefont {{Nakamura}}},
  \bibinfo {author} {\bibfnamefont {Q.}~\bibnamefont {{Lu}}}, \bibinfo {author}
  {\bibfnamefont {A.}~\bibnamefont {{Du}}}, \bibinfo {author} {\bibfnamefont
  {T.}~\bibnamefont {{Zhang}}}, \bibinfo {author} {\bibfnamefont
  {W.}~\bibnamefont {{Baumjohann}}}, \bibinfo {author} {\bibfnamefont {Y.~V.}\
  \bibnamefont {{Khotyaintsev}}}, \bibinfo {author} {\bibfnamefont
  {M.}~\bibnamefont {{Volwerk}}}, \bibinfo {author} {\bibfnamefont
  {M.}~\bibnamefont {{Andr{\'e}}}}, \bibinfo {author} {\bibfnamefont
  {M.}~\bibnamefont {{Fujimoto}}}, \bibinfo {author} {\bibfnamefont {T.~K.~M.}\
  \bibnamefont {{Nakamura}}}, \bibinfo {author} {\bibfnamefont {A.~N.}\
  \bibnamefont {{Fazakerley}}}, \bibinfo {author} {\bibfnamefont
  {J.}~\bibnamefont {{Du}}}, \bibinfo {author} {\bibfnamefont {W.}~\bibnamefont
  {{Teh}}}, \bibinfo {author} {\bibfnamefont {E.~V.}\ \bibnamefont {{Panov}}},
  \bibinfo {author} {\bibfnamefont {B.}~\bibnamefont {{Zieger}}}, \bibinfo
  {author} {\bibfnamefont {Y.}~\bibnamefont {{Pan}}}, \ and\ \bibinfo {author}
  {\bibfnamefont {S.}~\bibnamefont {{Lu}}},\ }\href {\doibase
  10.1029/2011JA017384} {\bibfield  {journal} {\bibinfo  {journal} {Journal of
  Geophysical Research (Space Physics)}\ }\textbf {\bibinfo {volume} {117}},\
  \bibinfo {eid} {A07223} (\bibinfo {year} {2012})}\BibitemShut {NoStop}%
\bibitem [{\citenamefont {Newman}\ \emph {et~al.}(2012)\citenamefont {Newman},
  \citenamefont {Goldman}, \citenamefont {Lapenta},\ and\ \citenamefont
  {Markidis}}]{Newman:2012}%
  \BibitemOpen
  \bibfield  {author} {\bibinfo {author} {\bibfnamefont {D.}~\bibnamefont
  {Newman}}, \bibinfo {author} {\bibfnamefont {M.}~\bibnamefont {Goldman}},
  \bibinfo {author} {\bibfnamefont {G.}~\bibnamefont {Lapenta}}, \ and\
  \bibinfo {author} {\bibfnamefont {S.}~\bibnamefont {Markidis}},\ }\href@noop
  {} {\bibfield  {journal} {\bibinfo  {journal} {Bulletin of the American
  Physical Society}\ }\textbf {\bibinfo {volume} {57}} (\bibinfo {year}
  {2012})}\BibitemShut {NoStop}%
\bibitem [{\citenamefont {Pritchett}(2007)}]{Pritchett:2007}%
  \BibitemOpen
  \bibfield  {author} {\bibinfo {author} {\bibfnamefont {P.~L.}\ \bibnamefont
  {Pritchett}},\ }\href@noop {} {\bibfield  {journal} {\bibinfo  {journal}
  {Physics of Plasmas}\ }\textbf {\bibinfo {volume} {14}},\ \bibinfo {pages}
  {052102} (\bibinfo {year} {2007})}\BibitemShut {NoStop}%
\bibitem [{\citenamefont {Fadeev}\ \emph {et~al.}(1965)\citenamefont {Fadeev},
  \citenamefont {Kvabtskhava},\ and\ \citenamefont {Komarov}}]{Fadeev:1965}%
  \BibitemOpen
  \bibfield  {author} {\bibinfo {author} {\bibfnamefont {V.}~\bibnamefont
  {Fadeev}}, \bibinfo {author} {\bibfnamefont {I.}~\bibnamefont {Kvabtskhava}},
  \ and\ \bibinfo {author} {\bibfnamefont {N.}~\bibnamefont {Komarov}},\
  }\href@noop {} {\bibfield  {journal} {\bibinfo  {journal} {Nuclear fusion}\
  }\textbf {\bibinfo {volume} {5}},\ \bibinfo {pages} {202} (\bibinfo {year}
  {1965})}\BibitemShut {NoStop}%
\bibitem [{\citenamefont {Markidis}\ \emph {et~al.}(2010)\citenamefont
  {Markidis}, \citenamefont {Lapenta},\ and\ \citenamefont
  {Rizwan-uddin}}]{Markidis:2010}%
  \BibitemOpen
  \bibfield  {author} {\bibinfo {author} {\bibfnamefont {S.}~\bibnamefont
  {Markidis}}, \bibinfo {author} {\bibfnamefont {G.}~\bibnamefont {Lapenta}}, \
  and\ \bibinfo {author} {\bibnamefont {Rizwan-uddin}},\ }\href@noop {}
  {\bibfield  {journal} {\bibinfo  {journal} {Mathematics and Computers in
  Simulation}\ }\textbf {\bibinfo {volume} {80}},\ \bibinfo {pages} {1509 }
  (\bibinfo {year} {2010})}\BibitemShut {NoStop}%
\bibitem [{\citenamefont {Goedbloed}\ and\ \citenamefont
  {Poedts}(2004)}]{Poedts:2004}%
  \BibitemOpen
  \bibfield  {author} {\bibinfo {author} {\bibfnamefont {J.~P.}\ \bibnamefont
  {Goedbloed}}\ and\ \bibinfo {author} {\bibfnamefont {S.}~\bibnamefont
  {Poedts}},\ }\href@noop {} {\emph {\bibinfo {title} {Principles of
  magnetohydrodynamics: With applications to laboratory and astrophysical
  plasmas}}}\ (\bibinfo  {publisher} {Cambridge University Press},\ \bibinfo
  {year} {2004})\BibitemShut {NoStop}%
\bibitem [{\citenamefont {Yoon}\ \emph {et~al.}(2002)\citenamefont {Yoon},
  \citenamefont {Lui},\ and\ \citenamefont {Sitnov}}]{yoon:2002}%
  \BibitemOpen
  \bibfield  {author} {\bibinfo {author} {\bibfnamefont {P.~H.}\ \bibnamefont
  {Yoon}}, \bibinfo {author} {\bibfnamefont {A.~T.}\ \bibnamefont {Lui}}, \
  and\ \bibinfo {author} {\bibfnamefont {M.~I.}\ \bibnamefont {Sitnov}},\
  }\href@noop {} {\bibfield  {journal} {\bibinfo  {journal} {Physics of
  Plasmas}\ }\textbf {\bibinfo {volume} {9}},\ \bibinfo {pages} {1526}
  (\bibinfo {year} {2002})}\BibitemShut {NoStop}%
\bibitem [{\citenamefont {Daughton}(2003)}]{Daughton:2003}%
  \BibitemOpen
  \bibfield  {author} {\bibinfo {author} {\bibfnamefont {W.}~\bibnamefont
  {Daughton}},\ }\href@noop {} {\bibfield  {journal} {\bibinfo  {journal}
  {Physics of Plasmas}\ }\textbf {\bibinfo {volume} {10}},\ \bibinfo {pages}
  {3103} (\bibinfo {year} {2003})}\BibitemShut {NoStop}%
\bibitem [{\citenamefont {Nagai}\ \emph {et~al.}(2001)\citenamefont {Nagai},
  \citenamefont {Shinohara}, \citenamefont {Fujimoto}, \citenamefont {Hoshino},
  \citenamefont {Saito}, \citenamefont {Machida},\ and\ \citenamefont
  {Mukai}}]{Nagai:2001}%
  \BibitemOpen
  \bibfield  {author} {\bibinfo {author} {\bibfnamefont {T.}~\bibnamefont
  {Nagai}}, \bibinfo {author} {\bibfnamefont {I.}~\bibnamefont {Shinohara}},
  \bibinfo {author} {\bibfnamefont {M.}~\bibnamefont {Fujimoto}}, \bibinfo
  {author} {\bibfnamefont {M.}~\bibnamefont {Hoshino}}, \bibinfo {author}
  {\bibfnamefont {Y.}~\bibnamefont {Saito}}, \bibinfo {author} {\bibfnamefont
  {S.}~\bibnamefont {Machida}}, \ and\ \bibinfo {author} {\bibfnamefont
  {T.}~\bibnamefont {Mukai}},\ }\href@noop {} {\bibfield  {journal} {\bibinfo
  {journal} {Journal of Geophysical Research: Space Physics}\ }\textbf
  {\bibinfo {volume} {106}},\ \bibinfo {pages} {25929} (\bibinfo {year}
  {2001})}\BibitemShut {NoStop}%
\bibitem [{\citenamefont {Karimabadi}\ \emph {et~al.}(2007)\citenamefont
  {Karimabadi}, \citenamefont {Daughton},\ and\ \citenamefont
  {Scudder}}]{karimabadi:2007}%
  \BibitemOpen
  \bibfield  {author} {\bibinfo {author} {\bibfnamefont {H.}~\bibnamefont
  {Karimabadi}}, \bibinfo {author} {\bibfnamefont {W.}~\bibnamefont
  {Daughton}}, \ and\ \bibinfo {author} {\bibfnamefont {J.}~\bibnamefont
  {Scudder}},\ }\href@noop {} {\bibfield  {journal} {\bibinfo  {journal}
  {Geophysical research letters}\ }\textbf {\bibinfo {volume} {34}},\ \bibinfo
  {pages} {L13104} (\bibinfo {year} {2007})}\BibitemShut {NoStop}%
\bibitem [{\citenamefont {Nagai}\ \emph {et~al.}(2013)\citenamefont {Nagai},
  \citenamefont {Shinohara}, \citenamefont {Zenitani}, \citenamefont
  {Nakamura}, \citenamefont {Nakamura}, \citenamefont {Fujimoto}, \citenamefont
  {Saito}, \citenamefont {Machida},\ and\ \citenamefont {Mukai}}]{Nagai:2013}%
  \BibitemOpen
  \bibfield  {author} {\bibinfo {author} {\bibfnamefont {T.}~\bibnamefont
  {Nagai}}, \bibinfo {author} {\bibfnamefont {I.}~\bibnamefont {Shinohara}},
  \bibinfo {author} {\bibfnamefont {S.}~\bibnamefont {Zenitani}}, \bibinfo
  {author} {\bibfnamefont {R.}~\bibnamefont {Nakamura}}, \bibinfo {author}
  {\bibfnamefont {T.~K.~M.}\ \bibnamefont {Nakamura}}, \bibinfo {author}
  {\bibfnamefont {M.}~\bibnamefont {Fujimoto}}, \bibinfo {author}
  {\bibfnamefont {Y.}~\bibnamefont {Saito}}, \bibinfo {author} {\bibfnamefont
  {S.}~\bibnamefont {Machida}}, \ and\ \bibinfo {author} {\bibfnamefont
  {T.}~\bibnamefont {Mukai}},\ }\href@noop {} {\bibfield  {journal} {\bibinfo
  {journal} {Journal of Geophysical Research: Space Physics}\ ,\ \bibinfo
  {pages} {accepted for publication}} (\bibinfo {year} {2013})}\BibitemShut
  {NoStop}%
\bibitem [{\citenamefont {Goldman}\ \emph {et~al.}(2011)\citenamefont
  {Goldman}, \citenamefont {Lapenta}, \citenamefont {Newman}, \citenamefont
  {Markidis},\ and\ \citenamefont {Che}}]{Goldman:2011}%
  \BibitemOpen
  \bibfield  {author} {\bibinfo {author} {\bibfnamefont {M.~V.}\ \bibnamefont
  {Goldman}}, \bibinfo {author} {\bibfnamefont {G.}~\bibnamefont {Lapenta}},
  \bibinfo {author} {\bibfnamefont {D.~L.}\ \bibnamefont {Newman}}, \bibinfo
  {author} {\bibfnamefont {S.}~\bibnamefont {Markidis}}, \ and\ \bibinfo
  {author} {\bibfnamefont {H.}~\bibnamefont {Che}},\ }\href@noop {} {\bibfield
  {journal} {\bibinfo  {journal} {Phys. Rev. Lett.}\ }\textbf {\bibinfo
  {volume} {107}},\ \bibinfo {pages} {135001} (\bibinfo {year}
  {2011})}\BibitemShut {NoStop}%
\bibitem [{\citenamefont {{Lapenta}}\ \emph {et~al.}(2010)\citenamefont
  {{Lapenta}}, \citenamefont {{Markidis}}, \citenamefont {{Divin}},
  \citenamefont {{Goldman}},\ and\ \citenamefont {{Newman}}}]{Lapenta:2010}%
  \BibitemOpen
  \bibfield  {author} {\bibinfo {author} {\bibfnamefont {G.}~\bibnamefont
  {{Lapenta}}}, \bibinfo {author} {\bibfnamefont {S.}~\bibnamefont
  {{Markidis}}}, \bibinfo {author} {\bibfnamefont {A.}~\bibnamefont {{Divin}}},
  \bibinfo {author} {\bibfnamefont {M.}~\bibnamefont {{Goldman}}}, \ and\
  \bibinfo {author} {\bibfnamefont {D.}~\bibnamefont {{Newman}}},\ }\href@noop
  {} {\bibfield  {journal} {\bibinfo  {journal} {Physics of Plasmas}\ }\textbf
  {\bibinfo {volume} {17}},\ \bibinfo {pages} {082106} (\bibinfo {year}
  {2010})}\BibitemShut {NoStop}%
\bibitem [{\citenamefont {Lapenta}\ \emph {et~al.}(2011)\citenamefont
  {Lapenta}, \citenamefont {Markidis}, \citenamefont {Divin}, \citenamefont
  {Goldman},\ and\ \citenamefont {Newman}}]{LapentaGRL:2011}%
  \BibitemOpen
  \bibfield  {author} {\bibinfo {author} {\bibfnamefont {G.}~\bibnamefont
  {Lapenta}}, \bibinfo {author} {\bibfnamefont {S.}~\bibnamefont {Markidis}},
  \bibinfo {author} {\bibfnamefont {A.}~\bibnamefont {Divin}}, \bibinfo
  {author} {\bibfnamefont {M.~V.}\ \bibnamefont {Goldman}}, \ and\ \bibinfo
  {author} {\bibfnamefont {D.~L.}\ \bibnamefont {Newman}},\ }\href@noop {}
  {\bibfield  {journal} {\bibinfo  {journal} {Geophys. Res. Lett.}\ }\textbf
  {\bibinfo {volume} {38}} (\bibinfo {year} {2011})}\BibitemShut {NoStop}%
\bibitem [{\citenamefont {Markidis}\ \emph {et~al.}(2011)\citenamefont
  {Markidis}, \citenamefont {Lapenta}, \citenamefont {Goldman}, \citenamefont
  {Newman},\ and\ \citenamefont {Andersson}}]{MarkidisPOP:2011}%
  \BibitemOpen
  \bibfield  {author} {\bibinfo {author} {\bibfnamefont {S.}~\bibnamefont
  {Markidis}}, \bibinfo {author} {\bibfnamefont {G.}~\bibnamefont {Lapenta}},
  \bibinfo {author} {\bibfnamefont {M.~V.}\ \bibnamefont {Goldman}}, \bibinfo
  {author} {\bibfnamefont {D.}~\bibnamefont {Newman}}, \ and\ \bibinfo {author}
  {\bibfnamefont {L.}~\bibnamefont {Andersson}},\ }\href@noop {} {\bibfield
  {journal} {\bibinfo  {journal} {submitted to Physics of Plasmas}\ } (\bibinfo
  {year} {2011})}\BibitemShut {NoStop}%
\bibitem [{\citenamefont {{Shibata}}\ and\ \citenamefont
  {{Tanuma}}(2001)}]{Shibata:2001}%
  \BibitemOpen
  \bibfield  {author} {\bibinfo {author} {\bibfnamefont {K.}~\bibnamefont
  {{Shibata}}}\ and\ \bibinfo {author} {\bibfnamefont {S.}~\bibnamefont
  {{Tanuma}}},\ }\href@noop {} {\bibfield  {journal} {\bibinfo  {journal}
  {Earth, Planets, and Space}\ }\textbf {\bibinfo {volume} {53}},\ \bibinfo
  {pages} {473} (\bibinfo {year} {2001})}\BibitemShut {NoStop}%
\bibitem [{\citenamefont {Lapenta}\ \emph {et~al.}(2003)\citenamefont
  {Lapenta}, \citenamefont {Brackbill},\ and\ \citenamefont
  {Daughton}}]{Lapenta:2003}%
  \BibitemOpen
  \bibfield  {author} {\bibinfo {author} {\bibfnamefont {G.}~\bibnamefont
  {Lapenta}}, \bibinfo {author} {\bibfnamefont {J.}~\bibnamefont {Brackbill}},
  \ and\ \bibinfo {author} {\bibfnamefont {W.}~\bibnamefont {Daughton}},\
  }\href@noop {} {\bibfield  {journal} {\bibinfo  {journal} {Physics of
  plasmas}\ }\textbf {\bibinfo {volume} {10}},\ \bibinfo {pages} {1577}
  (\bibinfo {year} {2003})}\BibitemShut {NoStop}%
\bibitem [{\citenamefont {{Karimabadi}}\ \emph {et~al.}(1999)\citenamefont
  {{Karimabadi}}, \citenamefont {{Krauss-Varban}}, \citenamefont {{Omidi}},\
  and\ \citenamefont {{Vu}}}]{Karimabadi:1999}%
  \BibitemOpen
  \bibfield  {author} {\bibinfo {author} {\bibfnamefont {H.}~\bibnamefont
  {{Karimabadi}}}, \bibinfo {author} {\bibfnamefont {D.}~\bibnamefont
  {{Krauss-Varban}}}, \bibinfo {author} {\bibfnamefont {N.}~\bibnamefont
  {{Omidi}}}, \ and\ \bibinfo {author} {\bibfnamefont {H.~X.}\ \bibnamefont
  {{Vu}}},\ }\href@noop {} {\bibfield  {journal} {\bibinfo  {journal} {Journal
  of Geophysical Research (Space Physics)}\ }\textbf {\bibinfo {volume}
  {1041}},\ \bibinfo {pages} {12313} (\bibinfo {year} {1999})}\BibitemShut
  {NoStop}%
\bibitem [{\citenamefont {{Drake}}\ \emph
  {et~al.}(2006{\natexlab{b}})\citenamefont {{Drake}}, \citenamefont
  {{Swisdak}}, \citenamefont {{Schoeffler}}, \citenamefont {{Rogers}},\ and\
  \citenamefont {{Kobayashi}}}]{Drake:2006}%
  \BibitemOpen
  \bibfield  {author} {\bibinfo {author} {\bibfnamefont {J.~F.}\ \bibnamefont
  {{Drake}}}, \bibinfo {author} {\bibfnamefont {M.}~\bibnamefont {{Swisdak}}},
  \bibinfo {author} {\bibfnamefont {K.~M.}\ \bibnamefont {{Schoeffler}}},
  \bibinfo {author} {\bibfnamefont {B.~N.}\ \bibnamefont {{Rogers}}}, \ and\
  \bibinfo {author} {\bibfnamefont {S.}~\bibnamefont {{Kobayashi}}},\
  }\href@noop {} {\bibfield  {journal} {\bibinfo  {journal} {Geophys. Res.
  Lett.}\ }\textbf {\bibinfo {volume} {33}},\ \bibinfo {pages} {L13105}
  (\bibinfo {year} {2006}{\natexlab{b}})}\BibitemShut {NoStop}%
\bibitem [{\citenamefont {{Retin{\`o}}}\ \emph {et~al.}(2008)\citenamefont
  {{Retin{\`o}}}, \citenamefont {{Nakamura}}, \citenamefont {{Vaivads}},
  \citenamefont {{Khotyaintsev}}, \citenamefont {{Hayakawa}}, \citenamefont
  {{Tanaka}}, \citenamefont {{Kasahara}}, \citenamefont {{Fujimoto}},
  \citenamefont {{Shinohara}}, \citenamefont {{Eastwood}}, \citenamefont
  {{Andr{\'e}}}, \citenamefont {{Baumjohann}}, \citenamefont {{Daly}},
  \citenamefont {{Kronberg}},\ and\ \citenamefont
  {{Cornilleau-Wehrlin}}}]{Retino:2008}%
  \BibitemOpen
  \bibfield  {author} {\bibinfo {author} {\bibfnamefont {A.}~\bibnamefont
  {{Retin{\`o}}}}, \bibinfo {author} {\bibfnamefont {R.}~\bibnamefont
  {{Nakamura}}}, \bibinfo {author} {\bibfnamefont {A.}~\bibnamefont
  {{Vaivads}}}, \bibinfo {author} {\bibfnamefont {Y.}~\bibnamefont
  {{Khotyaintsev}}}, \bibinfo {author} {\bibfnamefont {T.}~\bibnamefont
  {{Hayakawa}}}, \bibinfo {author} {\bibfnamefont {K.}~\bibnamefont
  {{Tanaka}}}, \bibinfo {author} {\bibfnamefont {S.}~\bibnamefont
  {{Kasahara}}}, \bibinfo {author} {\bibfnamefont {M.}~\bibnamefont
  {{Fujimoto}}}, \bibinfo {author} {\bibfnamefont {I.}~\bibnamefont
  {{Shinohara}}}, \bibinfo {author} {\bibfnamefont {J.~P.}\ \bibnamefont
  {{Eastwood}}}, \bibinfo {author} {\bibfnamefont {M.}~\bibnamefont
  {{Andr{\'e}}}}, \bibinfo {author} {\bibfnamefont {W.}~\bibnamefont
  {{Baumjohann}}}, \bibinfo {author} {\bibfnamefont {P.~W.}\ \bibnamefont
  {{Daly}}}, \bibinfo {author} {\bibfnamefont {E.~A.}\ \bibnamefont
  {{Kronberg}}}, \ and\ \bibinfo {author} {\bibfnamefont {N.}~\bibnamefont
  {{Cornilleau-Wehrlin}}},\ }\href {\doibase 10.1029/2008JA013511} {\bibfield
  {journal} {\bibinfo  {journal} {Journal of Geophysical Research (Space
  Physics)}\ }\textbf {\bibinfo {volume} {113}},\ \bibinfo {eid} {A12215}
  (\bibinfo {year} {2008})}\BibitemShut {NoStop}%
\bibitem [{\citenamefont {{Wang}}\ \emph {et~al.}(2010)\citenamefont {{Wang}},
  \citenamefont {{Lu}}, \citenamefont {{Li}}, \citenamefont {{Huang}},\ and\
  \citenamefont {{Wang}}}]{Wang:2010}%
  \BibitemOpen
  \bibfield  {author} {\bibinfo {author} {\bibfnamefont {R.}~\bibnamefont
  {{Wang}}}, \bibinfo {author} {\bibfnamefont {Q.}~\bibnamefont {{Lu}}},
  \bibinfo {author} {\bibfnamefont {X.}~\bibnamefont {{Li}}}, \bibinfo {author}
  {\bibfnamefont {C.}~\bibnamefont {{Huang}}}, \ and\ \bibinfo {author}
  {\bibfnamefont {S.}~\bibnamefont {{Wang}}},\ }\href {\doibase
  10.1029/2010JA015473} {\bibfield  {journal} {\bibinfo  {journal} {Journal of
  Geophysical Research (Space Physics)}\ }\textbf {\bibinfo {volume} {115}},\
  \bibinfo {eid} {A11201} (\bibinfo {year} {2010})}\BibitemShut {NoStop}%
\end{thebibliography}
\providecommand{\noopsort}[1]{}\providecommand{\singleletter}[1]{#1}%

\end{document}